\documentclass[aps,preprint,groupedaddress,a4paper]{revtex4-2}

\usepackage[T1]{fontenc}   
\usepackage{textcomp} 
\usepackage{graphicx}

\usepackage{mathtools}
\usepackage[colorlinks = true,linkcolor = blue,urlcolor  = blue,citecolor = blue,anchorcolor = blue]{hyperref}
\usepackage{xcolor, soul}
\usepackage{color}

\usepackage{makecell}

\DeclarePairedDelimiter\ket{\lvert}{\rangle}
\DeclarePairedDelimiter\bra{\langle}{\rvert}

\begin{document}


\title{Designing quantum error correction codes for practical spin qudit systems}


\author{Sumin Lim}
\affiliation{CAESR, Department of Physics, University of Oxford, The Clarendon Laboratory, Parks Road, Oxford OX1 3PU, UK}
\author{Arzhang Ardavan}
\affiliation{CAESR, Department of Physics, University of Oxford, The Clarendon Laboratory, Parks Road, Oxford OX1 3PU, UK}


\date{\today}

\begin{abstract}
The implementation of practical error correction protocols is essential for deployment of quantum information technologies. Ways of exploiting high-spin nuclei, which have multi-level quantum resources, have attracted interest in this context because they offer additional Hilbert space dimensions in a spatially compact and theoretically efficient structure. We present a quantitative analysis of the performance of a spin-qudit-based error-correctable quantum memory, with reference to the actual Hamiltonians of several potential candidate systems. First, the ideal code-word implemented on a spin-7/2 nucleus, which provides first order Pauli-$X$, $Y$ and $Z$ error correction, has intrinsic infidelity due to mixed eigenstates under realistic conditions. We confirm that expansion to a spin-9/2 system with tailored code-words can compensate this infidelity. Second, we claim that electric field fluctuations -- which are inevitable in real systems -- should  also be considered as a noise source, and we illustrate an encoding/decoding scheme for a multi-spin-qudit-based error correction code that can simultaneously compensate for both electric and magnetic field perturbations. Such strategies are important as we move beyond the current noisy-intermediate quantum era, and fidelities above two or three nines becomes crucial for implementation of quantum technologies. 
\end{abstract}


\maketitle

\section{Introduction}

In the past decade, quantum information processing strategies deploying high-dimensional quantum objects have gained significant attention from researchers across a range of disciplines. This has well-known advantages in quantum error correction (QEC), for example in the Gottesman-Kitaev-Preskill (GKP) code~\cite{gottesman2001encoding, pirandola2008minimal, noh2020encoding}, which has been studied extensively on both theoretical~\cite{albert2020robust, royer2020stabilization} and experimental levels~\cite{hu2019quantum,fluhmann2020direct, ofek2016extending, campagne2020quantum}. This approach can provide a spatially compact structure compared to qubit-based correction schemes\cite{gottesman1997stabilizer, knill1997theory, knill2000theory, google2023suppressing, terhal2015quantum, divincenzo1996fault, laflamme1996perfect}.
In this context, high spin nuclei offer promising candidates for encoding fault-tolerant logical qubits~\cite{chiesa2020molecular, gross2021designing}, not only owing to their superior coherence times, but also because their potential multi-dimensional spin subspace yields a hardware-efficient structure. Moreover, several recent reports claim that spin-qudit-based information processing can outperform same level of qubit-based counterparts in terms of resources~\cite{lim2023fault} and gate~\cite{jankovic2024noisy} efficiency. Thus, various theoretical~\cite{lockyer2021targeting, carretta2021perspective, omanakuttan2023multispin, omanakuttan2024fault, omanakuttan2023spin, omanakuttan2023qudit} and experimental~\cite{lim2025demonstrating, yu2025schrodinger} studies investigate ways of exploiting these systems as memory units for quantum technology.  For instance: nuclei from 3d transition metals -- such as Cu, V, Mn, hosting nuclear spins of 3/2 or 5/2 –- can provide correction of Pauli-$Z$ error; larger nuclei –- such as Sb, Bi, La, which have spins of 7/2 or 9/2 –- can provide simultaneous correction of all Pauli-$X$, $Y$, and $Z$ errors. 

Practical implementation of these ideas, however, comes with several challenges arising from the actual physical interaction terms in the Hamiltonian, which are usually unavoidable in condensed matter systems. In the existing literature, the spin state has typically been modelled based on eigenstates of the $S_Z$ operator, with first- or second-order errors described as compositions of the ideal spin operators, $S_X$, $S_Y$ and $S_Z$. In all cases, a sufficiently large quantising magnetic field $B_0$ will result in the states $\ket{m_S}$ and $\ket{m_I}$ being good eigenstates. However, but practical experiments are performed in finite field where this limit might not be reached, and therefore there is demand for quantitative analysis regarding errors originating from this issue. We find that existing QEC protocols with ideal code-word designs are not suitable for actual spin system with finite magnetic field, owing to perturbations such as hyperfine interaction, nuclear quadrupole interaction, crystal field splitting, and other terms, which lead to violations of the Knill-Laflamme (KL) criteria~\cite{knill1997theory}. 

So far, since we are at the noisy-intermediate-scale-quantum (NISQ) era~\cite{preskill2018quantum}, control and readout fidelities of systems lie around two nines or lower\cite{noiri2022fast, xue2022quantum, yoneda2018quantum}, therefore consideration for these weak perturbations has not been necessary. However, to go beyond the NISQ era and build a scalable quantum computer, we indeed need to tailor or reconstruct our code-word to make our quantum error correction protocols valid under these various higher-order terms. In the case of qubit-based QEC protocols, such as the surface code~\cite{arute2019quantum, wu2021strong, google2023suppressing, google2021exponential, bluvstein2024logical}, similar approaches -- tailoring code-words and decoders -- have been reported~\cite{bonilla2021xzzx, tuckett2020fault, tuckett2019tailoring, xu2023tailored} to compensate for biased error models. Moreover, optimisation of a spin qudit code-word~\cite{petiziol2021counteracting} to counteract $S_Z$ dephasing has also been suggested, based on numerical simulation of spin-bath interaction.  
We believe these approaches can be generalized; since we can parameterise the full spin Hamiltonian of a single quantum object, the effects of a range of errors can be investigated at a fundamental level, and effective code-word designs can be found for each system.

In this report, we provide a quantitative analysis regarding these effects. First, the existence of any non-Ising-type interaction (e.g., the isotropic hyperfine interaction) leads to mixing of $\ket{m_S,m_I}$ eigenstates in the $Z$-basis. 
We must take this mixing into account when tailoring code-words. 
Interestingly, a spin-7/2 system, which has been suggested as the minimal spin dimension for correcting first-order $S_X$, $S_Y$ and $S_Z$ errors, cannot host a code-word meeting the KL criteria in the presence of such an interaction. As we show in Section~\ref{tailored-code-word}, spin-9/2 system with tailored coefficients is required to deal with additional terms and conditions. Second, electron and nuclear spins with $m_S, m_I \ge 1$ are often exhibit electric-field-sensitive Hamiltonian terms, as confirmed experimentally in donor spins in semiconductors~\cite{fernandez2024navigating, asaad2020coherent} and nitrogen-vacancy centers in diamond~\cite{kim2015decoherence, jamonneau2016competition}.
Since, as is well known, magnons or phonons in condensed matter systems can cause magnetic or electric field fluctuations, realistic first-order error protection requires code-words robust against not only $B_X$, $B_Y$ and $B_Z$ fluctuations, but also $E_X$, $E_Y$ and $E_Z$ fluctuations. 
In Section~\ref{EandB-fluctuations}, we illustrate the construction of logical qubit code-words providing this protection, encoded in multiple nuclear spin qudits, as suggested conceptually in Refs~\cite{lim2023fault,  omanakuttan2023multispin}, with a working example of three coupled nuclear spins-7/2. To the best of our knowledge, this is the first report of a multi-spin-qudit encoding and decoding scheme. We focus on a model environment of group V donors in Si with magnetic field on the scale of 1~Tesla, but the approach we use here can be applied to a wide range of quantum high-spin systems in condensed matter.

\section{Tailored code-word with spin Hamiltonian}
\label{tailored-code-word}

One essential step in a quantum error correction protocol is, from a thermodynamic point of view, transferring extra entropy of the logical qubit into an ancillary qubit, and then transferring it into a thermal bath by initializing the ancillary qubit. Therefore, there must be coupling between the data qubit and the ancillary qubit, to perform, repeatedly, the entire cycle. 

In practice, the coupling between spin systems (e.g.\ the hyperfine interaction between nuclear and electron spins) generates an asymmetry in the spin Hamiltonian. For example, the Hamiltonian of the $S=1/2$, $I=7/2$ $^{123}$Sb donor spin in Si is, to a good approximation,
\begin{equation}
H = g \mu_B B \cdot S + g_I \mu_B B \cdot I + S \cdot A \cdot I \: 
\label{Si_Sb_H}
\end{equation}
where $g \mu_B= 28.02$~GHz/T, $g_I \mu_B=5.55$~MHz/T, and $A=101.52$~MHz (obtained from Ref.~\cite{morello2020donor}).

\begin{figure}
\includegraphics[width=16cm]{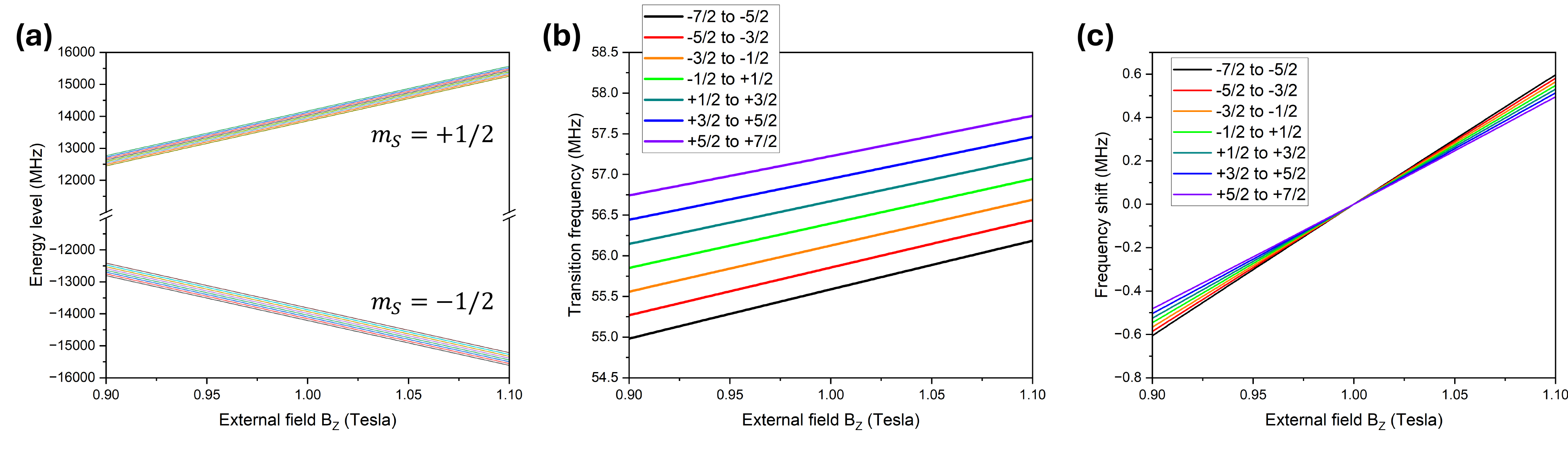}
\caption{(a) A energy level diagram for the spin states of the Si:Sb system.  (b) the nuclear spin transition frequencies within the $m_s = -1/2 $ manifold. (c) the shift of nuclear transition frequencies from their values at a field of 1~T as a function of magnetic field.
\label{energylevel}}
\end{figure}

Fig.~\ref{energylevel}(a) shows energy levels of this system, with an external magnetic field $B_Z$ around 1 Tesla applied along $Z$. Focussing on the lower energy group of eigenstates, those within the electron spin $m_S=-1/2$ manifold, Fig.~\ref{energylevel}(b) shows the seven $\Delta m_I = \pm 1$ transition frequencies between the nuclear spin states.  To show more clearly the magnetic field dependence of these frequencies, Fig.~1(c) shows the variation of the nuclear transition frequencies compared to their values at a magnetic field of 1~T.
As we can see in Fig.~\ref{energylevel}(c), the gradients of these transition frequencies against $B_Z$ differ from each other, reflecting the fact that the states $\ket{m_S, m_I}$ are not the exact eigenstates of the system owing to the off-diagonal terms in the isotropic hyperfine interaction term. The consequence of this is that a fluctuation $\Delta B_Z$ generates phase shifts on these levels differing from what is expected in the ideal cases described in, for example, Refs.~\cite{lim2023fault, gross2021designing}.We want also to note that $\ket{m_I}$ states mentioned hereafter are not perfect eigenstates of $I_Z$, but those for which there is a dominant component in the superposition.

\begin{figure}
\includegraphics[width=8cm]{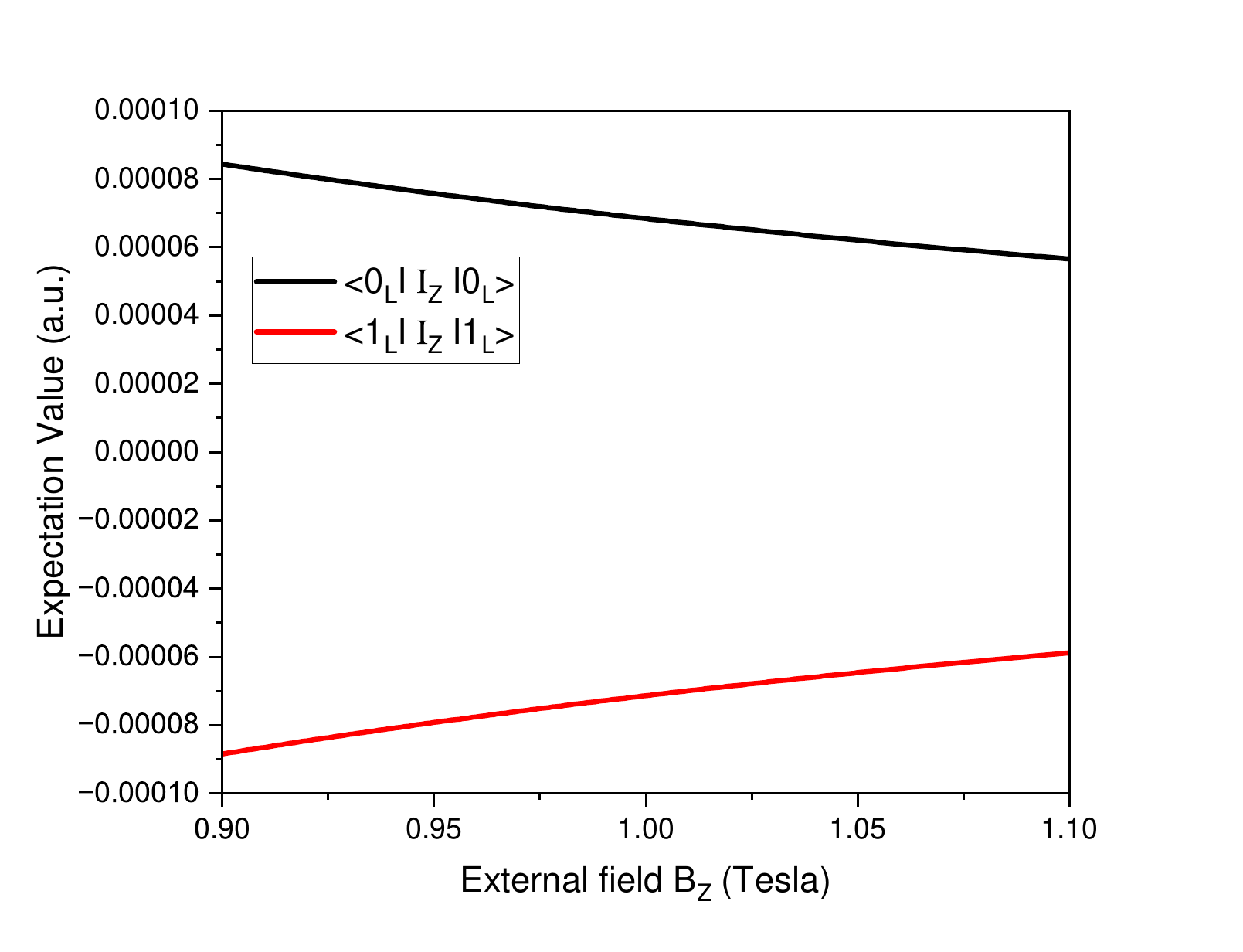}
\caption{ The expectation value for $\bra{0_L}I_Z\ket{0_L}$ and $\bra{1_L}I_Z\ket{1_L}$ for original 7/2 code-words, with practical environment of Si:Sb.
\label{Original_seven_half_z_projection}}
\end{figure}

Under these circumstances, the ideal logical code-words, $\ket{0_L}$ and $\ket{1_L}$, which are designed to protect against fluctuations of $B_Z$, $B_X$ and $B_Y$ under the assumption that only the nuclear spin Zeeman term plays a role, and are given by \cite{lim2023fault, gross2021designing} 
\begin{equation}
\begin{split}
\ket{0_L} = \sqrt{\frac{3}{10}} \ket*{-\frac{7}{2}} + \sqrt{\frac{7}{10}}\ket*{+\frac{3}{2}} \\
\ket{1_L} = -\sqrt{\frac{3}{10}} \ket*{+\frac{7}{2}} + \sqrt{\frac{7}{10}}\ket*{-\frac{3}{2}} \: ,
\end{split}
\label{Original_seven_half}
\end{equation}
do not satisfy the KL criteria under the influence of a fluctuation $\Delta B_Z$. This is demonstrated in Fig.~\ref{Original_seven_half_z_projection}, which shows the matrix elements $\bra{0_L}I_Z\ket{0_L}$  and $\bra{1_L}I_Z\ket{1_L}$ for magnetic fields around 1~T.  Both are non-zero, and furthermore, they differ from each other, leading to a violation of the KL criteria~\cite{knill1997theory},
\begin{equation}
\begin{split}
\bra{0_L} A^{\dagger}_{i}A_{j} \ket{1_L} =0 , \\
\bra{0_L} A^{\dagger}_{i}A_{j} \ket{0_L}  - \bra{1_L} A^{\dagger}_{i}A_{j} \ket{1_L} = 0 ,
\label{KL_criteria_eq}
\end{split}
\end{equation}
where $\ket{0_L}$ and $\ket{1_L}$ are the code words and $A_{i}$ represent the errors we want to protect against, i.e., the identity, and fluctuations of $B_X$, $B_Y$ and $B_Z$.

\begin{figure}
\includegraphics[width=8cm]{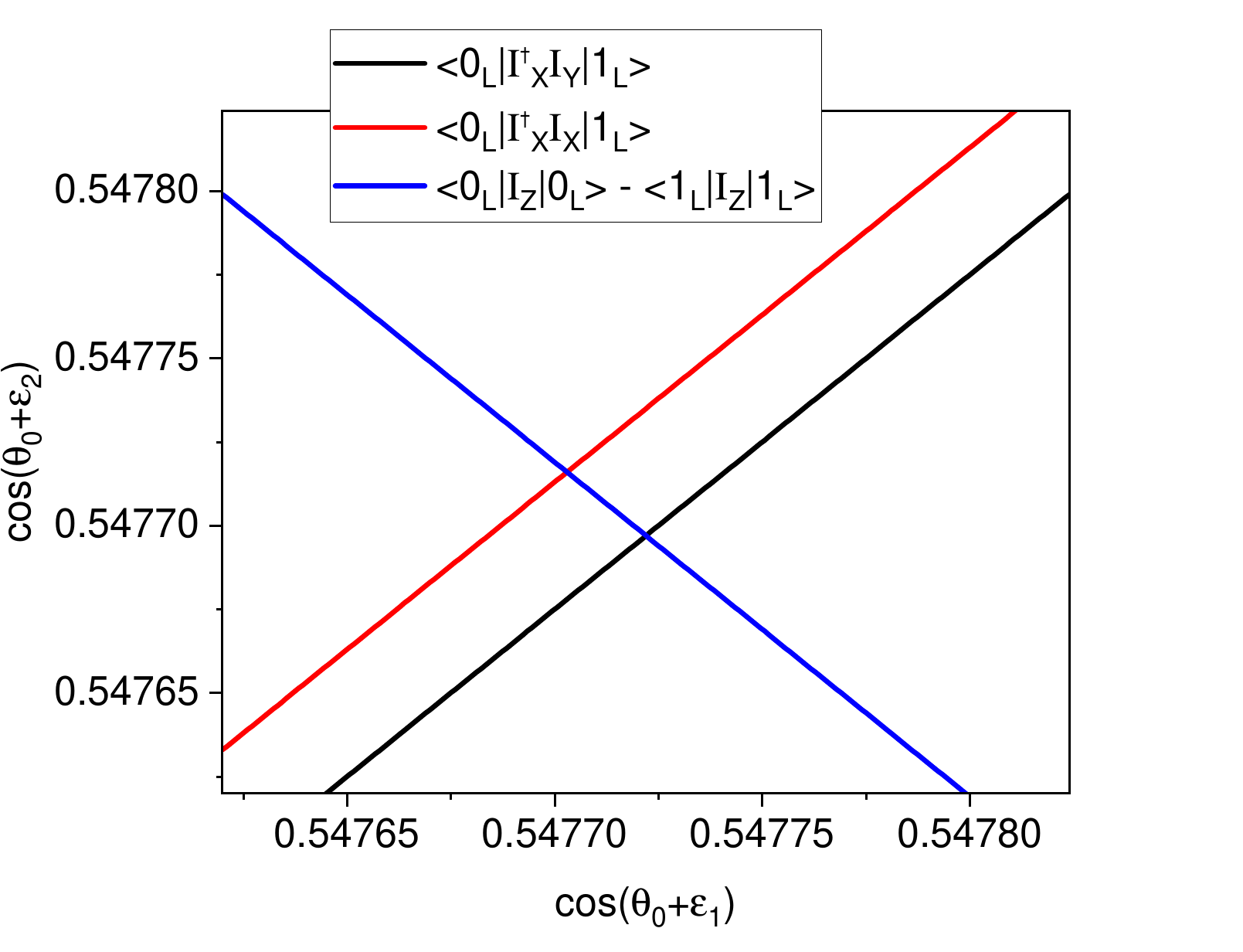}
\caption{ The lines show set of $\epsilon_1$ and $\epsilon_2$, which satisfying zero expectation value for $\bra{0_L}I_Z\ket{0_L} - \bra{1_L}I_Z\ket{1_L}$ (blue) , $\bra{0_L}I^{\dagger}_X I_X \ket{1_L}$ and (red) $\bra{0_L} I^{\dagger}_X I_Y \ket{1_L}$ (black). 
\label{Seven_half_Contour_plot}}
\end{figure}

This issue can be partially resolved if we also add an asymmetry in the code-word. For example, we can encode to protect against $\Delta B_Z$ perturbations by introducing the small distortion $( \epsilon_1 , \epsilon_2 )$ to the original code-word, such that
\begin{equation}
\begin{split}
\ket{0_L} = \cos(\theta_0+\epsilon_1) \ket*{-\frac{7}{2}} +  \sin(\theta_0+\epsilon_1)\ket*{+\frac{3}{2}} \\
\ket{1_L} = -\cos(\theta_0+\epsilon_2) \ket*{+\frac{7}{2}} + \sin(\theta_0+\epsilon_2)\ket*{-\frac{3}{2}} .
\end{split}
\end{equation}
Here, $ \theta_0 = \cos^{-1} \left( \sqrt{\frac{3}{10}} \right)$ and $ \epsilon_1 = \epsilon_2 = 0$ corresponds to the ideal case. Through careful choice of the two free parameters, $\epsilon_1$ and $\epsilon_2$, we can satisfy the condition $\bra{0_L}I_Z\ket{0_L}$ = $\bra{1_L}I_Z\ket{1_L}$. The blue line in Fig.~\ref{Seven_half_Contour_plot} shows a set of $( \epsilon_1, \epsilon_2 )$ that satisfying $\bra{0_L}I_Z\ket{0_L} - \bra{1_L}I_Z\ket{1_L}=0$ , demonstrating that this quantity is zero along a line in $( \epsilon_1, \epsilon_2 )$ space.

However, in this spin-7/2 case, we cannot also satisfy the equivalent KL criteria for fluctuations $\Delta B_X$ and $\Delta B_Y$ through a choice of $\epsilon_1$ and $\epsilon_2$:
the red and black lines in Fig.~\ref{Seven_half_Contour_plot} show the set of $( \epsilon_1 , \epsilon_2 )$ which have zero expectation value for $\bra{0_L}I^{\dagger}_X I_X \ket{1_L}$ and $\bra{0_L} I^{\dagger}_X I_Y \ket{1_L}$, respectively. As is evident from Fig.~\ref{Seven_half_Contour_plot}, there is no choice of $( \epsilon_1 , \epsilon_2 )$ for which all three quantities are zero, and therefore there is no simple way to identify logical qubits resilient to all three perturbations.

However, extending to a spin-9/2 system, as offered for example by  Bi dopants in Si, we acquire the resources required to deal with this asymmetry. As before, the isotropic hyperfine coupling of $I=9/2$ and $S=1/2$ generates mixing of $\ket{m_S, m_I}$ states, and an asymmetry in gradients of nuclear transition frequencies with magnetic field. 

We model Bi dopants in Si using the same Hamiltonian as Sb dopants, Eqn.~\ref{Si_Sb_H}, but with $S=1/2$, $I=9/2$, $g \mu_B= 28.02$~GHz/T, $g_I \mu_B=6.841$~MHz/T, and $A=1475.4$ MHz  (also obtained from Ref.~\cite{morello2020donor}). 
In this case, we can compensate for $\Delta B_X, \Delta B_Y,$ and $ \Delta B_Z$ fluctuations simultaneously, using the asymmetric code word
\begin{equation}
\begin{split}
\ket{0_L} = \cos(\theta_0+\epsilon_1) \ket*{-\frac{9}{2}} +  \sin(\theta_0+\epsilon_1)\ket*{+\frac{3}{2}} \\
\ket{1_L} = \cos(\theta_0+\epsilon_2) \ket*{+\frac{9}{2}} + \sin(\theta_0+\epsilon_2)\ket*{-\frac{3}{2}} .
\label{Nine_half_codeword}
\end{split}
\end{equation}
The case with $\theta_0 = \cos^{-1} \left( \frac{1}{2} \right)$ and  $ \epsilon_1 = \epsilon_2 = 0$  yields the original code-word for the spin-9/2 QEC code \cite{lim2023fault}. To this, we can add the distortion ($\epsilon_1, \epsilon_2$), to match KL criteria to account for the presence of the hyperfine term in the Hamiltonian. 

\begin{figure}
\includegraphics[width=8cm]{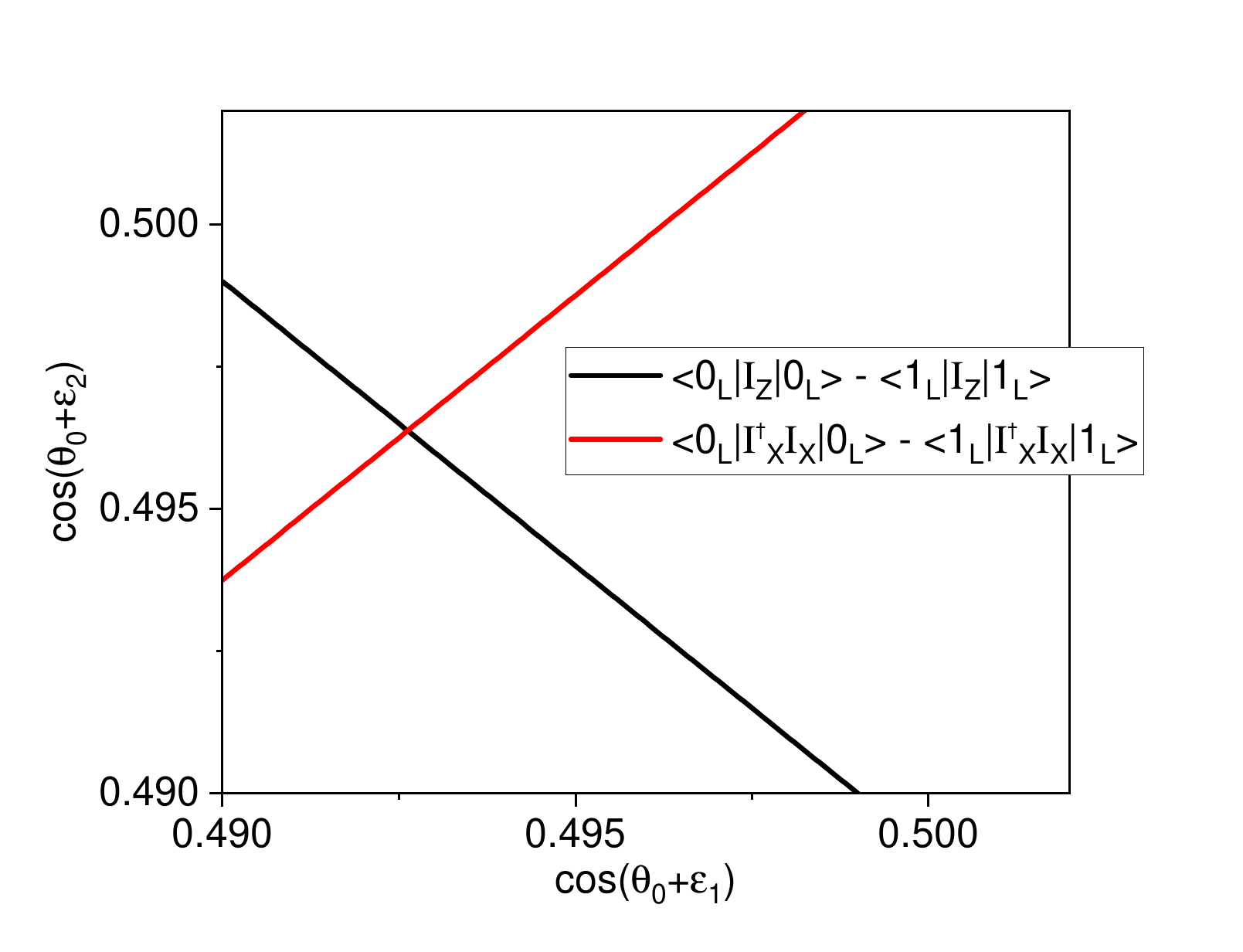}
\caption{ The lines show set of $\epsilon_1$ and $\epsilon_2$, which satisfying zero expectation value for $\bra{0_L}I_Z\ket{0_L} - \bra{1_L}I_Z\ket{1_L}$ (black) , $\bra{0_L}I^{\dagger}_X I_X \ket{0_L} -  \bra{1_L}I^{\dagger}_X I_X \ket{1_L}$ (red) , with ideal 9/2 code-word and additional $\epsilon_1$ and $\epsilon_2$ parameters. 
\label{Nine_half_Contour_plot}}
\end{figure}

The black and red lines in Fig.\ref{Nine_half_Contour_plot} shows set of  $( \epsilon_1 , \epsilon_2 )$, which have zero expectation value for $\bra{0_L}I_Z\ket{0_L} - \bra{1_L}I_Z\ket{1_L}$   , $\bra{0_L}I^{\dagger}_X I_X\ket{0_L} - \bra{1_L} I^{\dagger}_X I_X \ket{1_L}$  for spin 9/2 logical code-word in Eqn.~\ref{Nine_half_codeword}, respectively. The set of $( \epsilon_1 , \epsilon_2 )$ given by the black line also yields $\bra{0_L} I^{\dagger}_X I_Y \ket{0_L} - \bra{1_L} I^{\dagger}_X I_Y \ket{1_L}$ and the set represented by the red line yields $\bra{0_L}I^{\dagger}_Y I_Y\ket{0_L} - \bra{1_L} I^{\dagger}_Y I_Y \ket{1_L}$ and $\bra{0_L}I^{\dagger}_Z I_Z\ket{0_L} - \bra{1_L} I^{\dagger}_Z I_Z \ket{1_L}$ (as can be verified by considering the matrix elements or by numerical calculation).
In this 9/2 case, owing to the additional $m_I$ states between those making up the logical qubit, the other expectation values such as $\bra{0_L}I^{\dagger}_X I_X \ket{1_L}$  or $\bra{0_L}I^{\dagger}_X I_Y \ket{1_L}$ or $\bra{0_L}I^{\dagger}_Y I_Y \ket{1_L}$ are trivial zero. Thus, Fig.~\ref{Nine_half_Contour_plot} shows the dependences required to identify how to satisfy the KL criteria.

There exists a unique numerical solution, corresponding to $(\epsilon_1, \epsilon_2) = (0.0085...,0.0042...)$ (in radians), satisfying the KL criteria,
\begin{equation}
\begin{split}
\ket{0_L} =0.492625 \ket*{-\frac{9}{2}} +  0.870242 \ket*{+\frac{3}{2}} \\
\ket{1_L} = 0.496375 \ket*{-\frac{3}{2}} + 0.868208 \ket*{+\frac{9}{2}} .
\label{Nine_half_tailored}
\end{split}
\end{equation}
While this choice of $(\epsilon_1, \epsilon_2)$, and therefore the  coefficients in Eqn.~\ref{Nine_half_tailored}, are the specific solution for an external field of 1 Tesla, numerical solutions for other cases can be found easily by analogy. 

\begin{figure}
\includegraphics[width=8cm]{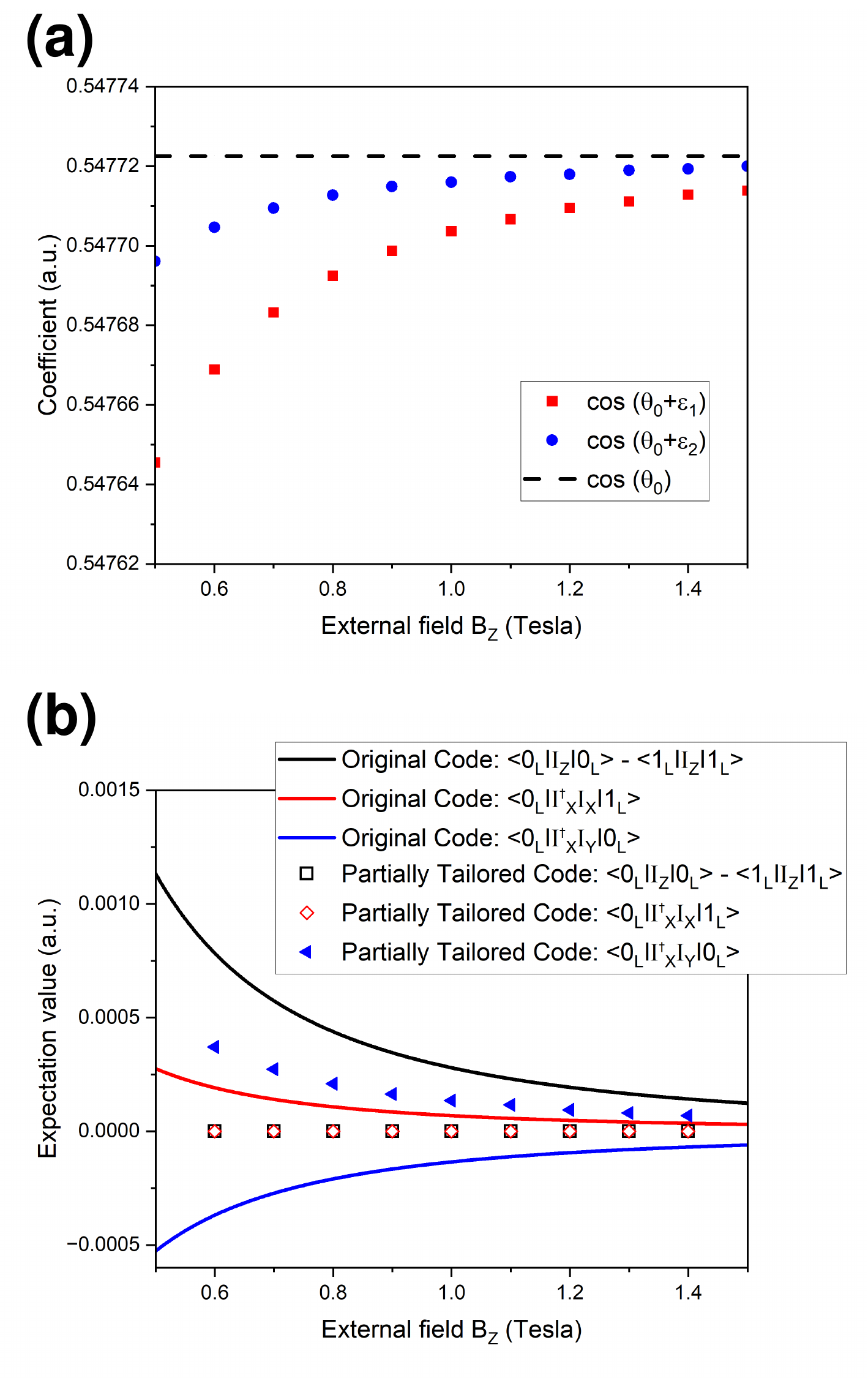}
\caption{ (a) $\epsilon_1 , \epsilon_2 $ for the partially tailored 7/2 code-words, which satisfies $\bra{0_L}I_Z\ket{0_L} - \bra{1_L}I_Z\ket{1_L} =0$ and $\bra{0_L}I^{\dagger}_X I_X\ket{1_L} = 0 $ conditions.  (b) The expectation values for KL criteria, for partially tailored code-words and original 7/2 code-words.
\label{Seven_partial_Tailored_codeword_graph}}
\end{figure}

In Fig.~\ref{Seven_partial_Tailored_codeword_graph}(a), we present the 
the amplitudes of $\ket{m_I}$ superposition components
for the partially tailored code-words for the spin-7/2 case, implemented for parameters for $^{123}$Sb:Si, as a function of magnetic field. These sets of $(\epsilon_1 , \epsilon_2)$ are solutions for $\bra{0_L}I_Z\ket{0_L} - \bra{1_L}I_Z\ket{1_L} =0$ and $\bra{0_L}I^{\dagger}_X I_X\ket{1_L} = 0$, but do not give $\bra{0_L} I^{\dagger}_X I_Y \ket{1_L} = 0$, because, as shown in Fig.~\ref{Seven_half_Contour_plot}, there is no global solution for all three conditions. We plot the deviation from satisfaction of the KL criteria for these partially tailored and original (untailored) spin-7/2 code words in Fig.~\ref{Seven_partial_Tailored_codeword_graph}(b), optimised as a function of magnetic field. As expected, the original spin-7/2 code word exhibits a finite deviation across the field range.
For these partially tailored cases, we can see that we can achive zero expectation values for $\bra{0_L}I_Z\ket{0_L} - \bra{1_L}I_Z\ket{1_L}$ and $\bra{0_L}I^{\dagger}_X I_X\ket{1_L}$ across the magnetic field range, there is always a finite imperfection for the $\bra{0_L} I^{\dagger}_X I_Y \ket{1_L}$ case.
An interesting feature is that the degree of imperfection decreases with increasing external field, and this is well matched with our intuition –- a larger external field brings eigenstates closer to ideal $S_Z$ eigenstates.

These 
deviations,
on the scale of $\sim10^{-3}$ to $\sim10^{-4}$, indicate the experimental circumstances under which it is important to consider the error arising from the hyperfine interaction. Some state-of-the-art experiments demonstrate control fidelities for single spin qubits and qudits around the three nines level \cite{xue2022quantum, yoneda2018quantum, fernandez2024navigating}, so this source of error is already experimentally significant.
The tailoring of the spin-7/2 code-word reduces these infidelities by a factor of roughly 3, but still there remains a finite amount of error because $S_X$ and $S_Y$ errors cannot be compensated fully. 

\begin{figure}
\includegraphics[width=8cm]{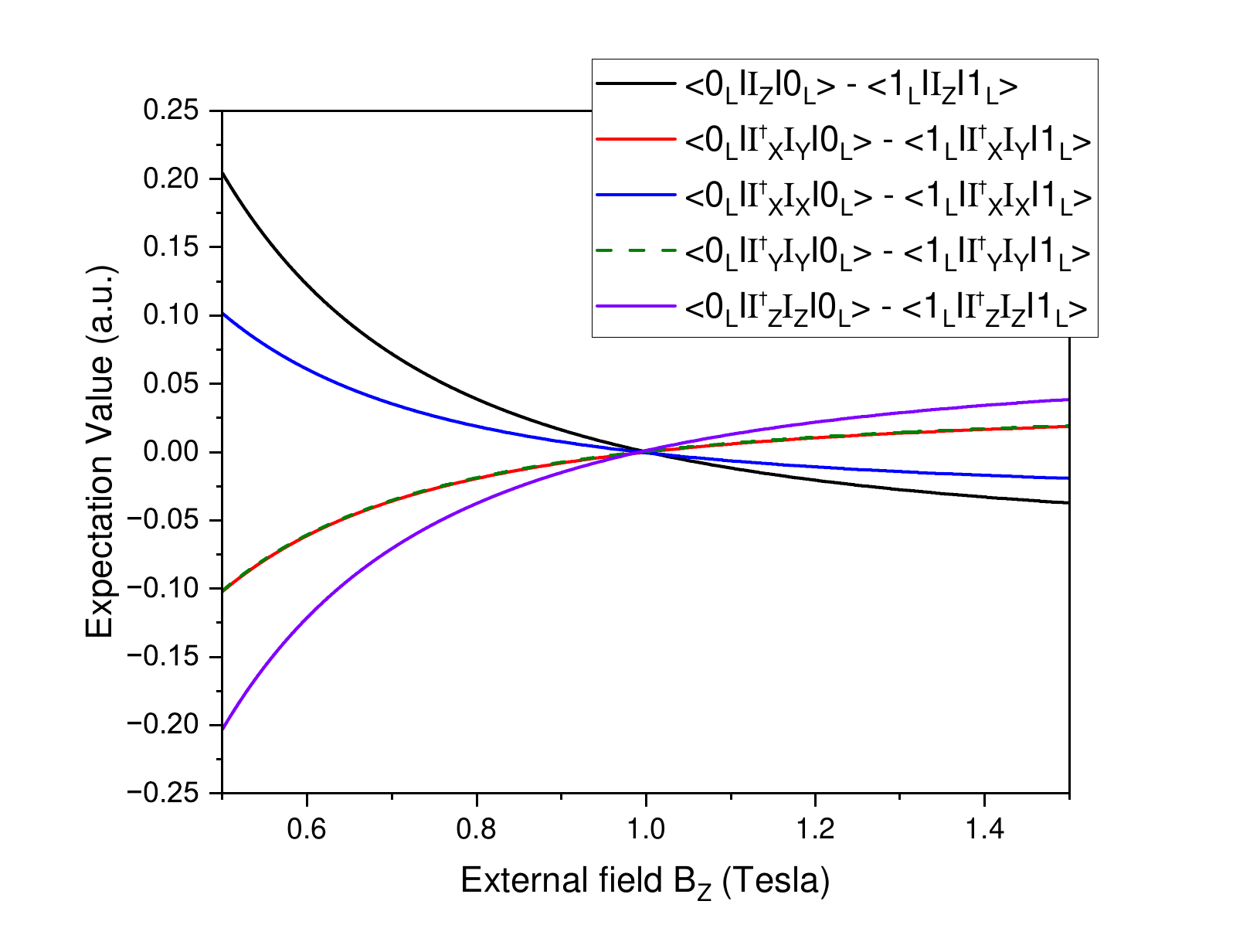}
\caption{ The infidelity for fully tailored 9/2 code-word for 1 Tesla coefficients case.
\label{Nine_Tailored_codeword_graph}}
\end{figure}

In Fig.~\ref{Nine_Tailored_codeword_graph}, we illustrate the deviations of the KL criteria in Eqn.~\ref{KL_criteria_eq} from 0 for the fully tailored spin-9/2 code-word (Eqn.~\ref{Nine_half_tailored}) implemented for parameters for Bi:Si at 1~Tesla, as a function of the applied external field strength. In this field regime, the spin-9/2 cases exhibit a larger error compared to the spin-7/2 case shown in Fig.~\ref{Seven_partial_Tailored_codeword_graph}(b), because of the large hyperfine coupling of Bi dopants ($\sim$ 1400 MHz) compared to Sb dopants ($\sim$ 100 MHz). However, as described above, the infidelity can be suppressed to exactly zero by tailoring the code-word for a particular magnetic field, 1 Tesla in this case.

Thus we have identified an approach to preparing robust quantum error-correctable memories using a nuclear spin-9/2, such as that offered by the Bi donor in Si, with a logical code-word tailored to account for the full spin Hamiltonian. Another practical way to avoid this problem is to use nuclear spin-7/2 without hyperfine coupling, as demonstrated in Ref.~\cite{asaad2020coherent}, storing the nucleus in the ionized state in which the donor electron is removed. Although the initial information encoding, detection (and correction) scheme requires coupling to an ancillary electron spin, the nuclear spin can be maintained in the ionized state during the memory storage time. This will minimize additional error due to asymmetry of the system Hamiltonian, depending on the ratio between memory storage time and the interaction time. 

\section{Logical qubit code-word protecting against electric and magnetic field fluctuations}
\label{EandB-fluctuations}

Often in solid state systems, the spin Hamiltonian can be coupled to an external electric field, particularly if the environment of the spin has broken inversion symmetry. Examples of Hamiltonian contributions through which electric field couplings can occur include, for nuclear spins, the nuclear quadrupole interaction $I \cdot Q \cdot I$ \cite{fernandez2024navigating, asaad2020coherent}, and for electron spins, crystal field terms such as $S \cdot D \cdot S$ and higher order terms \cite{george2013coherent, dolde2014nanoscale, dolde2011electric}.
(We note here that for other qudit candidates, such as superconducting circuits and trapped ions, practical implementations often rely on analogous higher-order Hamiltonian terms, to lift the degeneracies of transitions and allow spectral addressability.) 
From the engineering point of view, these spin-electric coupling terms are convenient, allowing coherent manipulation of spin states using electric control fields~\cite{fernandez2024navigating, asaad2020coherent, liu2019electric, liu2021coherent, liu2021quantum}.
However, from the perspective of information integrity and fault-tolerant encoding, these properties also imply that environmental noise of both electric and magnetic fields can disrupt quantum states. Indeed, electric field noise is known as one of the major limiting factors for spin coherence of nitrogen-vacancy centers~\cite{kim2015decoherence, jamonneau2016competition}.
By analogy with magnetic-field-fluctuation-related relaxation mechanisms such as thermal fluctuations of neighbouring spins (including environmental nuclear spin bath, and other electron spins via the dipolar interaction), scattering of phonons or charge fluctuations can induce electric field fluctuations at the site of spins~\cite{takahashi2011decoherence}. 

This suggests that the first order operators that we should protect against are not simply $S_X, S_Y$ and $S_Z$, but instead those associated with both magnetic and electric field perturbations.
Magnetic field perturbations generate, to first order, fluctuations of operators $S_X, S_Y$ and $S_Z$; experiments on spin-electric couplings in condensed matter~\cite{fernandez2024navigating, asaad2020coherent,  liu2019electric, liu2021coherent, liu2021quantum} suggest that we should also consider, to first order, operators of the form $S_X^2, S_Y^2, S_Z^2, S_X S_Y, S_Y S_Z$, and $S_Z S_X$ arising from electric field perturbations.

One way to make a logical code-word robust to all of these error operators
is by exploiting a single spin qudit dimension higher than 24 (i.e.,  $S \ge 23/2$). While second-order error correction of qubit-based code-word require minimum 11 qubits \cite{gottesman1996class} (i.e., a Hilbert space dimension of 2048) this 24-dimensional spin qudit can provide efficient encoding, and can provide robustness under any perturbation, because all significant practical perturbations on a spin can be described as fluctuation of either electric field or magnetic field.

In terms of experimental candidates for implementation, careful engineering of a giant single molecule magnet has the potential to build a collective electron spin of S=23/2 \cite{zabala2021single, murugesu2008large, takahashi2009coherent}. In this case, the spin Hamiltonian is described by a Zeeman term and a crystal field splitting term, and the logical codeword against first order $\vec{E}$ and $\vec{B}$ fluctuations can be prepared by analogy with the procedure described in Section~\ref{tailored-code-word}. Similarly, the code-word suggested in Ref.~\cite{lim2023fault} will be able to correct also for second order perturbations arising from fluctuations of $\vec{E}$ and $\vec{B}$:
\begin{equation}
\begin{split}
\ket{0_L} = +\sqrt{\frac{125}{1482}} \ket*{-\frac{23}{2}} +   \sqrt{\frac{874}{1482}} \ket*{-\frac{5}{2}} + \sqrt{\frac{483}{1482}} \ket*{+\frac{15}{2}}\\
\ket{1_L} = - \sqrt{\frac{125}{1482}} \ket*{+\frac{23}{2}} +   \sqrt{\frac{874}{1482}} \ket*{+\frac{5}{2}} + \sqrt{\frac{483}{1482}} \ket*{-\frac{15}{2}}\
\end{split}
\end{equation}

However, coherent and high-fidelity control of such a giant electron spin  remains difficult. Additionally, in general, electron spins have shorter relaxation times compared to nuclear spins. Therefore we propose another scheme to implement error correctable encoding, by using three coupled nuclear qudits of spin-7/2.

A logical code word satisfying the KL criteria for these error operators is
\begin{equation}
\begin{split}
\ket{0_L} = +\sqrt{\frac{2}{16}} \ket*{-\frac{7}{2}}_{A,B,C} +   \sqrt{\frac{7}{16}} \ket*{-\frac{3}{2}}_{A,B,C} + \sqrt{\frac{7}{16}} \ket*{+\frac{5}{2}}_{A,B,C}\\
\ket{1_L} = +\sqrt{\frac{2}{16}} \ket*{+\frac{7}{2}}_{A,B,C} +   \sqrt{\frac{7}{16}} \ket*{+\frac{3}{2}}_{A,B,C} - \sqrt{\frac{7}{16}} \ket*{-\frac{5}{2}}_{A,B,C}\
\label{Multi_qudit_code}
\end{split}
\end{equation}
where $ \ket{m_I}_{A,B,C}$ corresponds to the tensor product of three nuclear spin states, $ \ket{m_I}_A \otimes \ket{m_I}_B \otimes \ket{m_I}_C$. (A similar codeword given in Ref.~\cite{lim2023fault}, which consists of four spins-7/2 with the same coefficients, was proposed to protect against two individual errors. Here, three spins-7/2 are sufficient because we are considering a single error event, but accounting for second order operators.) Suitable corresponding physical systems may be prepared, for example, by precise ion implantation in semiconductors incorporating control circuitry~\cite{morello2020donor, tosi2017silicon, jamieson2017deterministic, groot2019deterministic}, or by chemical engineering of molecular magnets, such as Sc$_3$@C$_{82}$\cite{rahmer2005w, liu2018qubit}. While these three spins-7/2 qudits generate a large Hilbert space of dimension 512, we believe that this configuration is worth considering because it can host a code protecting against all realistic electromagnetic perturbations; for individual memory elements, further complexity up is unlikely to be necessary for practical technologies.

\begin{figure}
\includegraphics[width=16cm]{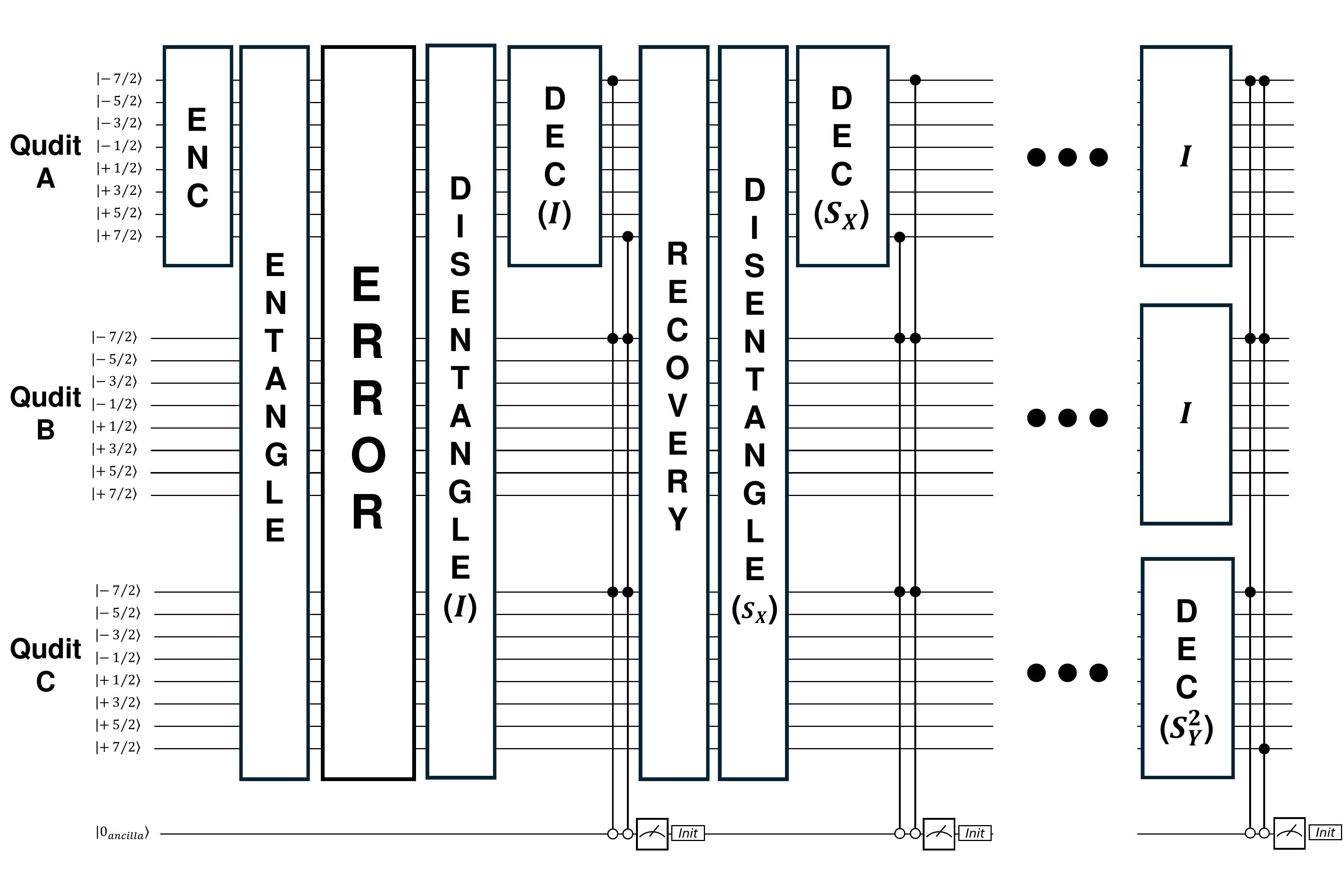}
\caption{ The high-level encoding and decoding scheme for the three spins-7/2 code word.
\label{Multi_qudit_code_diagram}}
\end{figure}

We now consider the detail of the sequence for the encoding and decoding of this multi-qudit error correction code, with the logical code word given in Eqn.~\ref{Multi_qudit_code}. For simplicity we assume that $\ket{m_I}_{A,B,C}$ are good eigenstates. A high-level schematic of the process is illustrated in Fig.~\ref{Multi_qudit_code_diagram}, with more detail on pulse sequences given in Appendices A and B. First, without loss of generality, we can start with the original qubit stored in a superposition of $\ket{-7/2},\ket{-5/2}$ of the first nuclear spin ($A$), in the form $\alpha\ket{-7/2}_A+\beta\ket{-5/2}_A$, while the other two nuclear spins ($B$ and $C$) are prepared in pure $\ket{-7/2}$ states. This starting state can be written as Eqn.~\ref{starting_state},

\begin{equation}
\begin{split}
\ket{\psi_0} = \left(\alpha \ket*{-\frac{7}{2}}_{A} + \beta \ket*{-\frac{5}{2}}_{A}\right) \otimes \ket*{-\frac{7}{2}}_B \otimes \ket*{-\frac{7}{2}}_C \:.
\label{starting_state}
\end{split}
\end{equation}

The logical qubit state is first encoded in qudit $A$ via the block of operations labeled ``ENC''. After this block, the state of the system is,
\begin{equation}
\begin{split}
\ket{\psi_1} = \left[ \alpha \left(+\sqrt{\frac{2}{16}} \ket*{-\frac{7}{2}}_{A} +   \sqrt{\frac{7}{16}} \ket*{-\frac{3}{2}}_{A} + \sqrt{\frac{7}{16}} \ket*{+\frac{5}{2}}_{A} \right) \right. \\
\left. +\beta \left( \sqrt{\frac{2}{16}} \ket*{+\frac{7}{2}}_{A} +   \sqrt{\frac{7}{16}} \ket*{+\frac{3}{2}}_{A} - \sqrt{\frac{7}{16}} \ket*{-\frac{5}{2}}_{A} \right) \right] \\
\otimes \ket*{-\frac{7}{2}}_{B} \otimes \ket*{-\frac{7}{2}}_{C} \:.
\label{State_after_enc}
\end{split}
\end{equation}

The qudits $B$ and $C$ are then entangled by operation block ``ENTANGLE'' (as described in more detail in Appendix~A), generating the state in Eqn.\ref{encoded_state}, with $\ket{0_L}$ and $\ket{1_L}$ defined in Eqn.~\ref{Multi_qudit_code},  
\begin{equation}
\ket{\psi_\mathrm{enc}} = \alpha \ket{0_L}+\beta \ket{1_L} \:.
\label{encoded_state}
\end{equation}

Following encoding, perturbations affect the system during the main storage time (block labelled ``ERROR''), after which the original information lies in multiple subspaces of the the three qudits. Since all error subspaces satisfy the KL criteria, it is possible to decode and detect each error case sequentially, one by one, as indicated by blocks labelled ``DISENTANGLE'', and ``DEC'' in Fig.~\ref{Multi_qudit_code_diagram}.

Note that since this error correction code is not pure~\cite{calderbank1998quantum} (or non-degenerate), this decoding scheme differs from the first-order single spin qudit error correction code in Eqn.~\ref{Original_seven_half}, which corresponds to a `pure' code. The number of error operators ($I, S_X, S_{X}^2, S_Y, \cdots, S_{Z}^2$) considered in this encoding is 12, bigger than dimension of the single spin-7/2 qudit space, 8. 
Therefore, there necessarily exists some mixing of logical zero and errored-logical zero (and also, logical one and errored-logical one). For example, because the $S_X$ operator is Hermitian, the projection of the original $\ket{0_L}$ to $S_X^2$ errors can be rewritten as $\bra{0_L}S_X^2\ket{0_L}=\bra{0_L}S_X^{\dagger}S_X\ket{0_L}=c_\mathrm{norm}$, where $c_\mathrm{norm}$ stands for a finite non-zero normalization factor.

We now provide more details of the actual decoding sequence. After applying an error operator on the logical qubit, the state can be written as
\begin{equation}
\begin{split}
E \ket{\psi_{enc}} =  \sqrt{1-\epsilon} (\alpha \ket{0_{L}}+ \beta \ket{1_{L}})      \\
+\sqrt{\epsilon_{X,A}}(\alpha S_{X,A}\ket{0_L}+\beta  S_{X,A} \ket{1_L}) \\
+\sqrt{\epsilon_{Y,A}}(\alpha  S_{Y,A}\ket{0_L}+\beta  S_{Y,A} \ket{1_L}) \\
+\sqrt{\epsilon_{X^2,A}}(\alpha S_{X,A}^2\ket{0_L}+\beta S_{X,A}^2 \ket{1_L}) \\
+\cdots \:, \\
\label{State_after_error}
\end{split}
\end{equation}
where $(\alpha \ket{0_L}+\beta \ket{1_L})$ represents the original information and $\epsilon_{i,j}$ indicates the probability of an $S_i$ error occurring on the qudit $j$. After this error, the ``DISENTANGLE($I$)'' and ``DEC($I$)'' gates can be expressed as the unitary operation $U_{D-I}$, which satisfies,
\begin{equation}
\begin{split}
U_{D-I}  \ket{0_{L}} = \ket{-7/2}_A \otimes \ket{-7/2}_B \otimes \ket{-7/2}_C     \\
U_{D-I}  \ket{1_{L}} = \ket{+7/2}_A \otimes \ket{-7/2}_B \otimes \ket{-7/2}_C  
\end{split}
\end{equation}

As shown in Appendix~B, this unitary transformation can be straightforwardly identified. The DISENTANGLE($I$) block is simply the inverse of the ENTANGLE block, and DEC($I$) is quite similar to the inverse of the ENC block. Then the conditional excitation from $\ket{-7/2}_A \otimes \ket{-7/2}_B \otimes \ket{-7/2}_C,  \ket{+7/2}_A \otimes \ket{-7/2}_B \otimes \ket{-7/2}_C$ to the electron spin ancilla is applied to detect the ``no error'' case. 

At this stage, we must check the effect of $U_{D-I}$ operator on other terms in the errored state, Eqn.~\ref{State_after_error}. The error terms such as $S_{X,A}\ket{0_L}$, $S_{X,A}\ket{1_L}$, $S_{Y,A}\ket{0_L}$, $S_{Y,A}\ket{1_L}$, $\cdots$ all lie in a  subspace orthogonal to both $ \ket{0_{L}}$ and $\ket{1_{L}}$. Therefore, they contribute zero to the coefficients of $\ket{-7/2}_A \otimes \ket{-7/2}_B \otimes \ket{-7/2}_C $  and $\ket{+7/2}_A \otimes \ket{-7/2}_B \otimes \ket{-7/2}_C$ following the $U_{D-I}$ operator, and are not affected by the conditional excitation to the ancilla.

In contrast, the error terms such as  $S_{X,A}^2\ket{0_L}$ , $S_{Y,A}^2\ket{0_L}$, $\cdots $ have nonzero inner product with original $\ket{0_{L}}$, and the equivalent is true for $\ket{1_{L}}$. As a working example, consider the effect of the $U_{D-I}$ operator on $S_{X,A}^2 (\alpha \ket{0_{L}} + \beta \ket{1_{L}})$; considering each logical qubit state in turn,
\begin{equation}
\begin{split}
U_{D-I}  S_{X,A}^2\ket{0_{L}} = U_{D-I} ( c_{XX}\ket{0_{L}} + (1-c_{XX})\ket{P_{X^{2},A,0}})     \\
= c_{XX}(\ket{-7/2}_A \otimes \ket{-7/2}_B \otimes \ket{-7/2}_C)\\
+(1-c_{XX})U_{D-I}\ket{P_{X^{2},A,0}} \\
U_{D-I}  S_{X,A}^2\ket{1_{L}} = U_{D-I} ( c_{XX}\ket{1_{L}} + (1-c_{XX})\ket{P_{X^{2},A,1}})     \\
= c_{XX}(\ket{+7/2}_A \otimes \ket{-7/2}_B \otimes \ket{-7/2}_C)\\
+(1-c_{XX})U_{D-I}\ket{P_{X^{2},A,1}} \\
\end{split}
\end{equation}
Here, we have separated $S_{X,A}^2\ket{0_{L}}$ into two orthogonal components, $\ket{0_{L}}$ and $\ket{P_{X^{2},A,0}}$ (and equivalent for $S_{X,A}^2\ket{1_{L}}$). Crucially, owing to the satisfaction of the KL criteria, $\bra{0_L}S_{X}^2\ket{0_L} = \bra{1_L}S_{X}^2\ket{1_L}$, the coefficient $c_{XX}$ is the same in both.

Therefore, after conditional excitation of the ancilla, and allowing for the possibility of occurrence of any of the correctable errors, the state becomes
\begin{equation}
\begin{split}
\ket{\Psi} = \mathrm{(remaining \, terms)} \ket{0_\mathrm{ancilla}}       \\
+\sqrt{1-\epsilon + c_{XX} + c_{XY} + \cdots } \left(\alpha \ket*{-\frac{7}{2}}_{A} \otimes \ket*{-\frac{7}{2}}_{B}  \otimes \ket*{-\frac{7}{2}}_{C} \right)\ket{1_\mathrm{ancilla}}  \\
+\sqrt{1-\epsilon + c_{XX} + c_{XY} + \cdots } \left(\beta  \ket*{+\frac{7}{2}}_{A} \otimes \ket*{-\frac{7}{2}}_{B}  \otimes \ket*{-\frac{7}{2}}_{C} \right)\ket{1_\mathrm{ancilla}}
\end{split}
\end{equation}
where $1-\epsilon$ corresponds to the probability of no error occurring, $c_{XX}$  corresponds to $\epsilon_{X^2}\bra{0_L}S_X^2\ket{0_L}$ , $c_{XY}$ corresponds to $\epsilon_{XY}\bra{0_L}S_XS_Y\ket{0_L}$,  and so on.

If a measurement of the ancilla yields ``1'', we conclude that an error did not occur and we can conclude the decoding process. Otherwise, we apply the ``RECOVERY'' block (which is the inverse of 
$U_{D-I}$). This restores the state to Eqn.~\ref{State_after_error} but with the 
un-errored component projected out, and with a corresponding change to the normalization factor. Thus we can iterate the decoding process over each error case, as shown in Fig.~\ref{Multi_qudit_code_diagram}. Note that when we detect
the error case $S_{X^2,A}$ (and other non-orthogonal cases), we need to decode for $\ket{P_{X^{2},A,0}}, \ket{P_{X^{2},A,1}}$ because the state has changed changed owing to the detection of the no-error case.

The entire decoding sequence is presented in Appendix B. One cycle of decoding requires roughly 1700 pulses. The typical relaxation times associated with $S_Z$ and $S_Z^2$ are related to the phase coherence time, $T_2$, of the system, and other terms, $S_X$, $S_Y$, $S_X^2$, $S_Y^2$, $\cdots$, are related to the longitudinal relaxation time, $T_1$. $T_1$ and $T_2$ are highly system dependent, but based on some assumptions we can simulate the efficiency this encoding.
Given that each round of the error correction sequence requires around 1700 pulses, we can estimate the how the overall fidelity depends on the practical single-pulse fidelities. 

Under the assumption that all first-order errors have same probability, to achieve advantage from this protocol, we require single-pulse fidelities better than $99.999 \%$ . We note that in solid state spin systems, $Z$-errors tend to occur frequently and are often the dominant source of decoherence. In that case, we can frequently check specific types of error (in this case, $S_Z, S_ZS_X, S_ZS_Y, S_Z^2$) on the decoding scheme, and effectively, reduce the required decoding scheme to about 500 pulses, thus improving the overall error-correction fidelity for the same level of single-pulse fidelity.

\section{Conclusion and future work}

We have demonstrated how to tailor fault-tolerant spin qudit code-words which can protect quantum information under realistic Hamiltonians in practical environments. The ideal code-word specified in Refs.~\cite{lim2023fault, gross2021designing} for the spin-7/2 case lies on somewhat unstable plateau originating from the 
strict spherical
symmetry; 
 even a single additional coupling term -- which must exist because an ancillary qubit is essential for QEC -- can destroy the symmetry, resulting in intrinsic infidelity. 
At the quantitative level, our numerical simulations provide baseline numbers for the maximum operational fidelity in example systems. For instance, a fault-tolerant code-word implementation on the nuclear spin-7/2 Si:Sb system with a typical X band or Q band EPR setup will suffer unavoidable errors of around $\sim 10^{-3}$ to $\sim 10^{-4}$ owing to the hyperfine coupling with the electron ancilla. These effects will become important when the fidelity of operations improves to a comparable level, and they should be considered as we move beyond the current NISQ era and achieve fault-tolerant information processing. 
Thanks to the availability of nuclei having spin-9/2, we confirm that the extra levels available can be exploited to compensate for perturbations from coupling terms. 

Another important characteristic of high-spin qudits is their interaction with electric fields owing to the quadrupole interaction. While magnetic field fluctuations generate first order errors, electric field fluctuations generate error operators that are second-order in the spin operators. In order to simultaneously protect against both electric and magnetic field fluctuations, we present the simplest form of a multi-qudit fault-tolerant encoding and decoding process, which can, in principle, be expanded into more general cases.

Depending on the specific physical system, specialized code-word designs for specific errors (e.g., a code-word for correction of $\Delta B_Z$ up to 3rd order, using a nuclear spin-7/2) can also be identified.  In the case of electric field induced noise, 
we can benefit from the recent demonstration of electric field control of nuclear spin qudits \cite{fernandez2024navigating, asaad2020coherent} to develop models of how to mitigate associated errors; more accurate models can be provided as more experimental exploration is conducted in the field. We believe this work motivates future studies on strategies that exploit nuclear spin qudits as basic building blocks of scalable quantum information processors.

\begin{acknowledgments}
This project was supported by the European Union's Horizon 2020 research and innovation programme under grant agreements 862893 (FATMOLS) and 863098 (SPRING).
\end{acknowledgments}


\bibliography{Fault_to}



\appendix

\setcounter{figure}{0}
\renewcommand{\figurename}{Fig.}
\renewcommand{\thefigure}{A\arabic{figure}}

\section{The schematic diagram of the multi-qudit-error correction encoding sequence}

\begin{figure}
\includegraphics[width=16cm]{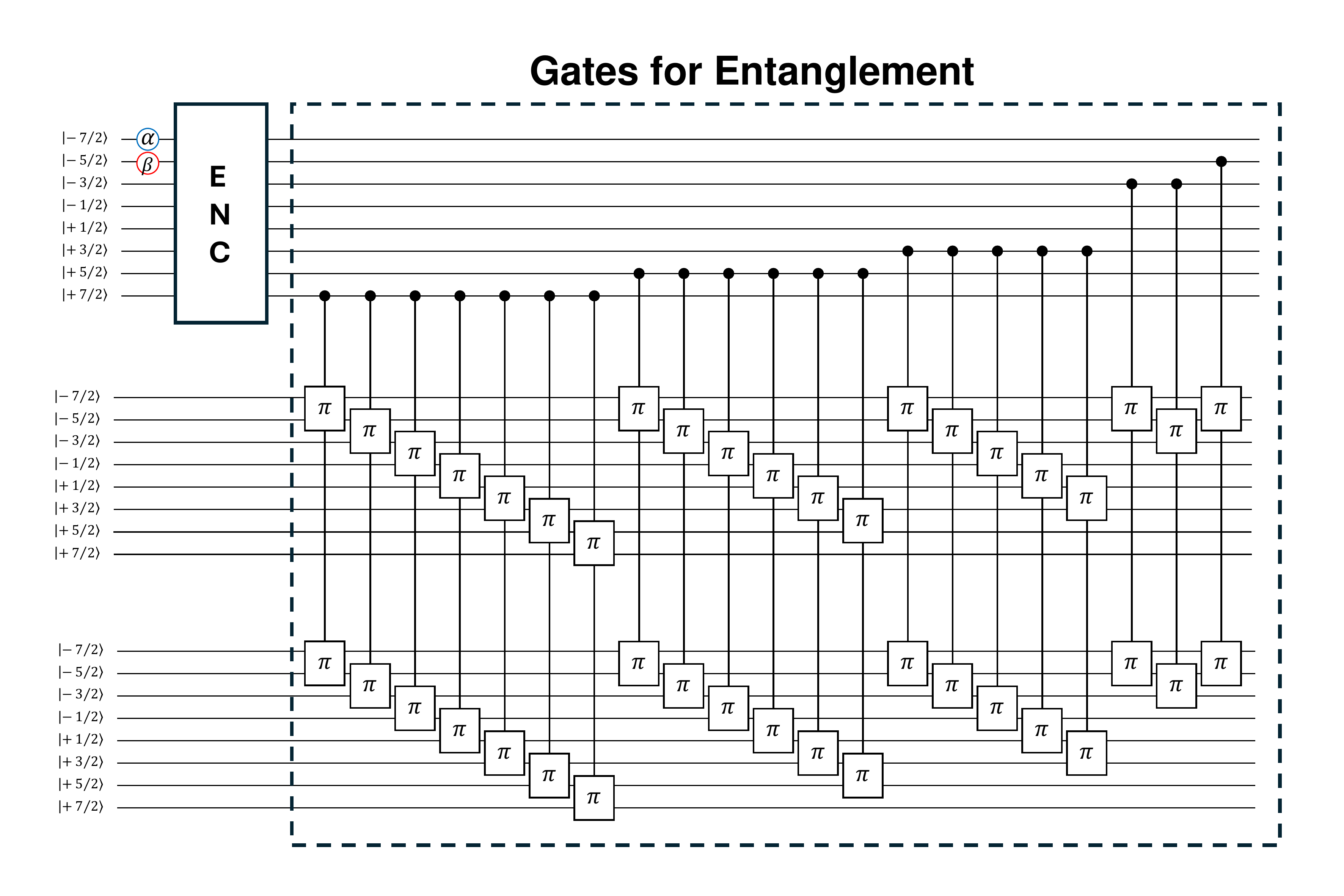}
\caption{ The schematic diagram of the full encoding pulse sequence for the multi-qudit error correction code. 
\label{Full-enc-sequence}}
\end{figure}

\begin{figure}
\includegraphics[width=16cm]{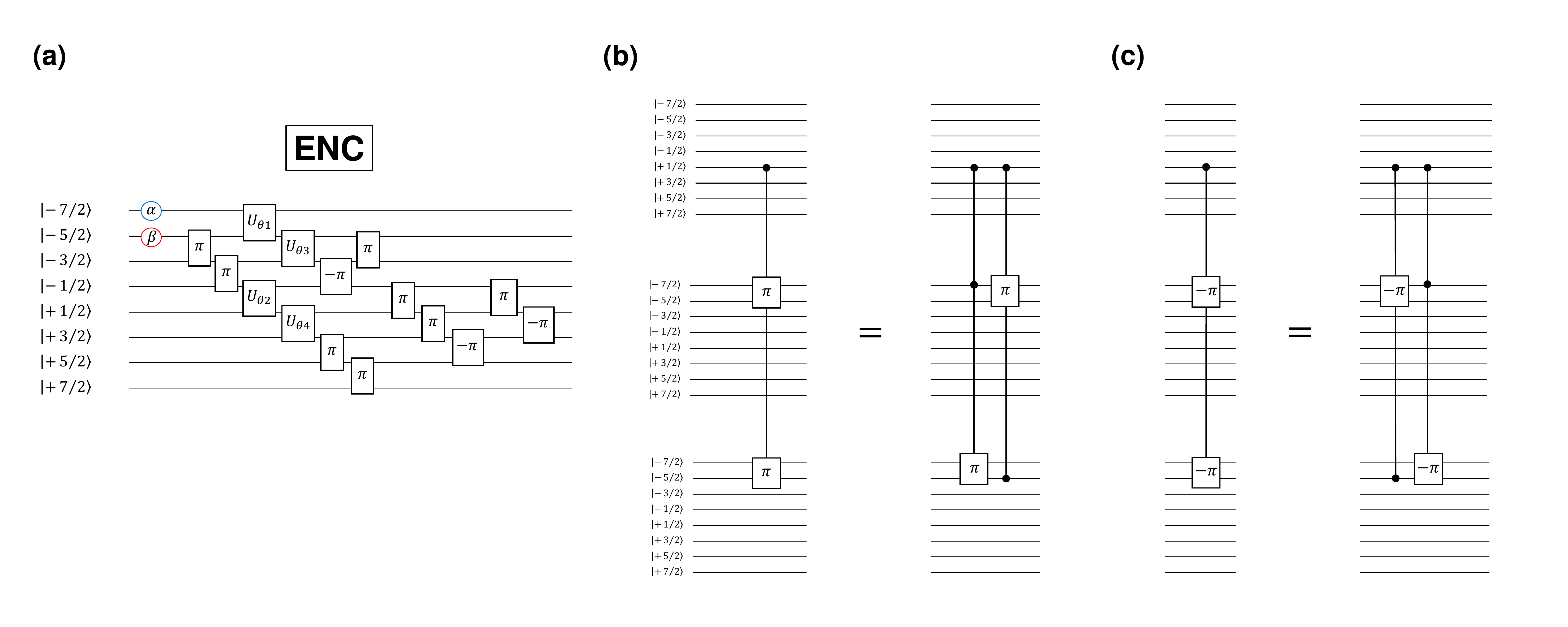}
\caption{ (a) The schematic diagram for the ENC pulse block. Unitary rotations $U_{\theta_i}$ are around the $y$ axis, with $\cos(\theta_i) = \sqrt{2/16}, \sqrt{7/16}, \sqrt{1/2}, \sqrt{7/9}$ for $i = 1, 2, 3, 4$, respectively. (b) The definition of controlled-double-$\pi$  (c) The definition of the inverse of controlled-double-$\pi$ . 
\label{ENC_gates_and_double_pi}}
\end{figure}

Fig.~\ref{Full-enc-sequence} shows full encoding sequence blocks for first-, second- and third qudit of multi-qudit-error-correction sequence. Without loss of generality, the starting state - original information on qudit A, pure $\ket{-7/2}$ preparation on qudit B and C - can be written in form of Eqn.~\ref{psi0}

\begin{equation}
\begin{split}
\ket{\psi_0} = \left( \alpha \ket*{-\frac{7}{2}}_{A} +   \beta \ket*{-\frac{5}{2}}_{A} \right) \otimes \ket*{-\frac{7}{2}}_B \otimes \ket*{-\frac{7}{2}}_C \:.
\label{psi0}
\end{split}
\end{equation}

From this general starting states, gates of ENC code block (as shown in Fig.~\ref{ENC_gates_and_double_pi} (a)) on qudit A will change state into Eqn.~\ref{psi1},

\begin{equation}
\begin{split}
\ket{\psi_1} = \left[ \alpha \left(+\sqrt{\frac{2}{16}} \ket*{-\frac{7}{2}}_{A} +   \sqrt{\frac{7}{16}} \ket*{-\frac{3}{2}}_{A} + \sqrt{\frac{7}{16}} \ket*{+\frac{5}{2}}_{A} \right) + \right. \\
 \left. \beta  \left(+\sqrt{\frac{2}{16}} \ket*{+\frac{7}{2}}_{A} +   \sqrt{\frac{7}{16}} \ket*{+\frac{3}{2}}_{A} - \sqrt{\frac{7}{16}} \ket*{-\frac{5}{2}}_{A} \right)  \right] \\
\otimes \ket*{-\frac{7}{2}}_B \otimes \ket*{-\frac{7}{2}}_C \:.
\label{psi1}
\end{split}
\end{equation}

After the ENC code block, conditional $\pi$ gates from qudit A to qudit B and C can generate entanglement of three qudits, as shown in Fig.~\ref{Full-enc-sequence}. (This is also labelled in ENTANGLE code block, in Fig.~\ref{Multi_qudit_code_diagram}.) In the diagram, controlled-dobule-$\pi$ gates are designed to apply controlled $\pi$ to the two other qudits, as defined in Fig.~\ref{ENC_gates_and_double_pi}(b). 
 Finally, the state will encoded into Eqn.~\ref{psi2}

\begin{equation}
\begin{split}
\ket{\psi_{enc}} = \alpha \left(+\sqrt{\frac{2}{16}} \ket*{-\frac{7}{2}}_{A,B,C} +   \sqrt{\frac{7}{16}} \ket*{-\frac{3}{2}}_{A,B,C} + \sqrt{\frac{7}{16}} \ket*{+\frac{5}{2}}_{A,B,C} \right) + \\
  \beta  \left( +\sqrt{\frac{2}{16}} \ket*{+\frac{7}{2}}_{A,B,C} +   \sqrt{\frac{7}{16}} \ket*{+\frac{3}{2}}_{A,B,C} - \sqrt{\frac{7}{16}} \ket*{-\frac{5}{2}}_{A,B,C} \right) \:.
\label{psi2}
\end{split}
\end{equation}

\section{The schematic diagram of multi-qudit-error correction decoding sequence}
\label{generalisation1}

In this Appendix, we  briefly introduce the detailed decoding sequences for the $I$ (``no-error'') case, the $S_X$ error on qudit A case, and the $S_X^2$ error on qudit A case. This provides the general framework to detect and correct other error cases.

\subsection{The decoding sequence for $I$ case }

\begin{figure}
\includegraphics[width=16cm]{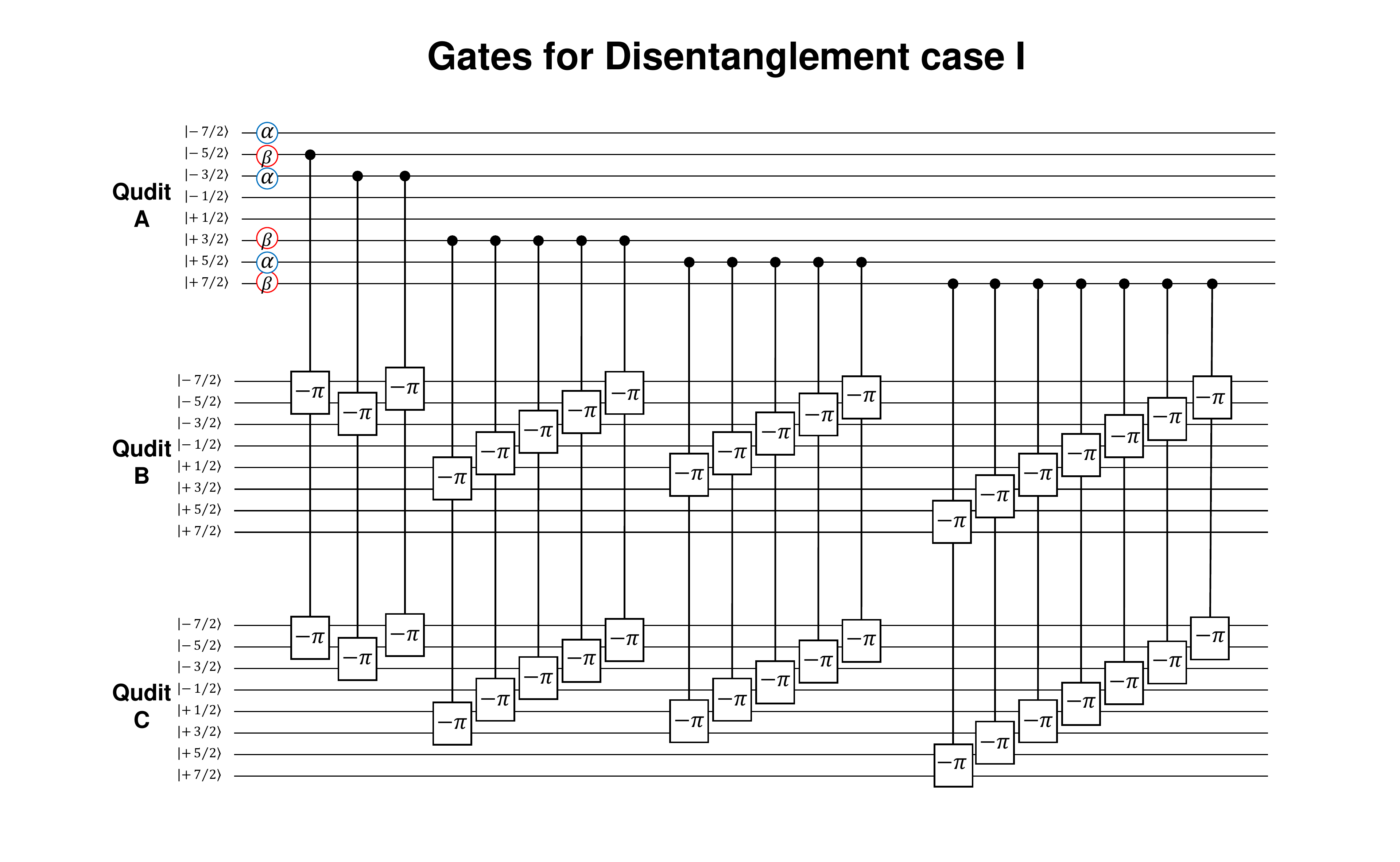}
\caption{ The Disentangle gates for I.
\label{Disentangle_I}}
\end{figure}

\begin{figure}
\includegraphics[width=8cm]{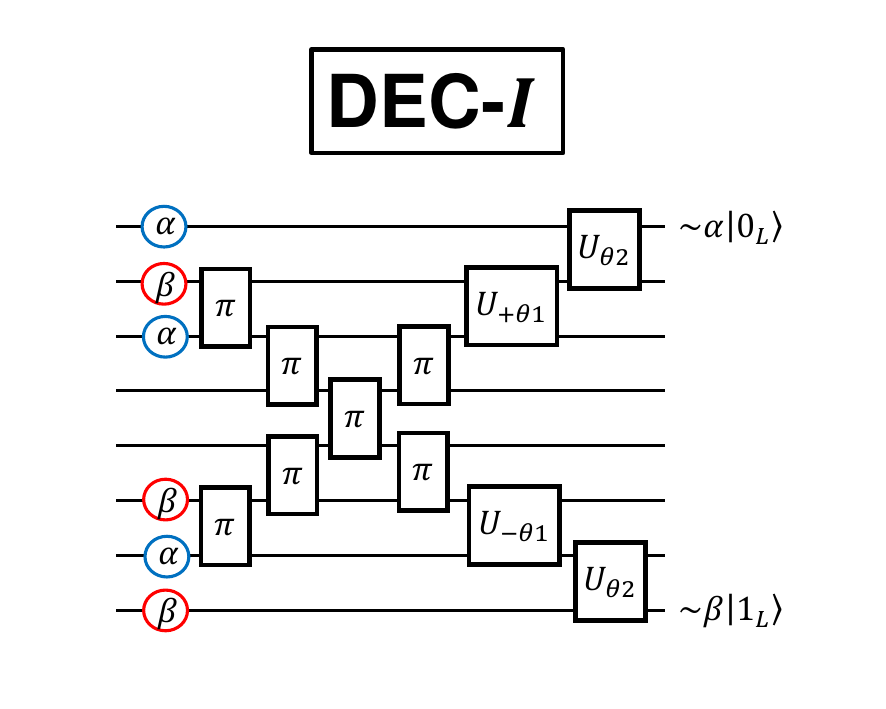}
\caption{ The DEC gates for I. Unitary rotations $U_{\theta_i}$ are around $y$ axis, with cos$(\theta_i) = \sqrt{1/2}, \sqrt{2/16}$ for $i = 1, 2$, respectively.
\label{DEC_I_gates}}
\end{figure}

After storage, our encoded state is represented by Eqn.~\ref{State_after_error}, and we will first apply the operation to detect the $I$ case. The DISENTANGLE($I$) gates, shown in Fig.~\ref{Disentangle_I} will transform the state into Eqn.~\ref{psi1}, because this is simply the inverse of the ENTANGLE circuit we applied during the encoding sequence.

Then, we can apply the DEC-$I$ operation in Fig.~\ref{DEC_I_gates}. This will send original $ \sqrt{1-\epsilon}  (\alpha \ket{0_{L}}+ \beta \ket{1_{L}})$ state into our target state, 
\begin{equation}
\begin{split}
\ket{\psi} =  \left( \sqrt{1-\epsilon}+C \right) \left( \alpha \ket*{-\frac{7}{2}}_{A} +   \beta \ket*{+\frac{7}{2}}_{A} \right) \otimes \ket*{-\frac{7}{2}}_B \otimes \ket*{-\frac{7}{2}}_C \\
+\mathrm{(remaining \, terms)} \:.
\label{after_dec_I}
\end{split}
\end{equation}

As discussed in main text, these gates correspond to the $U_{D-I}$ operation, and there can be additional coefficient arising from non-orthogonal terms. However, the target state still remains  separable, and therefore we can make a conditional excitation of the ancilla and finally measure (and detect) the ``no-error'' case. If the ancilla yields a ``0'' measurement, which means some kinds of error occured, we  move on to next step. We first apply the RECOVERY operation (Fig.~\ref{Multi_qudit_code_diagram}), which is defined as the inverse of $U_{D-I}$, taking the state will back to a form similar to Eqn.~\ref{State_after_error}, but without no error term and with new normalization factor.

\subsection{The decoding sequence for $S_X$ error on qudit A case }

Then, we can start the decoding of the $\sqrt{\epsilon_{X,A}}(\alpha S_{X,A}\ket{0_L}+\beta  S_{X,A} \ket{1_L})$ term. This state can be written as Eqn.~\ref{SX_on_enc},
\begin{equation}
\begin{split}
\ket{\psi_{S_X,A}} = \alpha \left(+\sqrt{\frac{2}{16}}\cdot\frac{\sqrt{7}}{2} \ket*{-\frac{5}{2}}_{A}\ket*{-\frac{7}{2}}_{B,C} +\sqrt{\frac{7}{16}}\cdot\frac{\sqrt{12}}{2} \ket*{-\frac{5}{2}}_{A}\ket*{-\frac{3}{2}}_{B,C} \right. \\
+\sqrt{\frac{7}{16}}\cdot\frac{\sqrt{15}}{2} \ket*{-\frac{1}{2}}_{A}\ket*{-\frac{3}{2}}_{B,C} 
+\sqrt{\frac{7}{16}}\cdot\frac{\sqrt{12}}{2} \ket*{+\frac{3}{2}}_{A}\ket*{+\frac{5}{2}}_{B,C} \\ 
\left. +\sqrt{\frac{7}{16}}\cdot\frac{\sqrt{7}}{2} \ket*{+\frac{7}{2}}_{A}\ket*{+\frac{5}{2}}_{B,C} \right) \\
+ \beta \left(+\sqrt{\frac{2}{16}}\cdot\frac{\sqrt{7}}{2} \ket*{+\frac{5}{2}}_{A}\ket*{+\frac{7}{2}}_{B,C} +\sqrt{\frac{7}{16}}\cdot\frac{\sqrt{12}}{2} \ket*{+\frac{5}{2}}_{A}\ket*{+\frac{3}{2}}_{B,C} \right. \\
+\sqrt{\frac{7}{16}}\cdot\frac{\sqrt{15}}{2} \ket*{+\frac{1}{2}}_{A}\ket*{+\frac{3}{2}}_{B,C} 
-\sqrt{\frac{7}{16}}\cdot\frac{\sqrt{12}}{2} \ket*{-\frac{3}{2}}_{A}\ket*{-\frac{5}{2}}_{B,C} \\
\left. -\sqrt{\frac{7}{16}}\cdot\frac{\sqrt{7}}{2} \ket*{-\frac{7}{2}}_{A}\ket*{-\frac{5}{2}}_{B,C} \right) \:.
 \label{SX_on_enc}
\end{split}
\end{equation}

\begin{figure}
\includegraphics[width=16cm]{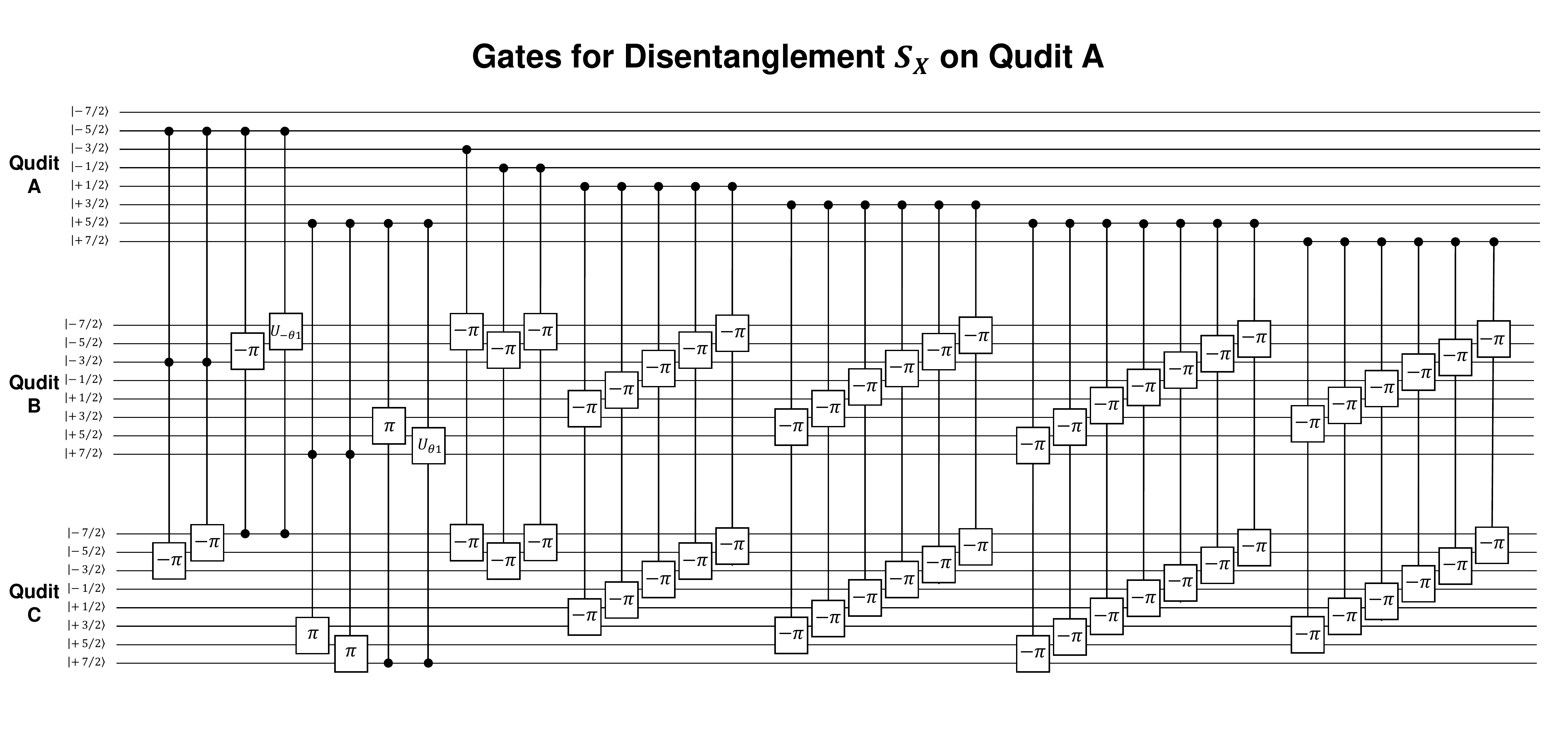}
\caption{ The DISENTANGLE($S_X$) operation on qudit A. The unitary rotation $U_{\theta_1}$ is around the $y$ axis, with $\cos(\theta_1) = \sqrt{1/7}$.
\label{Disentangle_SX}}
\end{figure}

On this state, we will apply the DISENTANGLE($S_X$) operation, as shown in Fig.~\ref{Disentangle_SX}. This will change the state into
\begin{equation}
\begin{split}
\ket{\psi_{S_X,A}} =  \left[ \alpha \left(+\frac{7\sqrt{2}}{8} \ket*{-\frac{5}{2}}_{A} 
+\frac{\sqrt{105}}{8} \ket*{-\frac{1}{2}}_{A} 
+\frac{\sqrt{84}}{8} \ket*{+\frac{3}{2}}_{A} 
+\frac{7}{8} \ket*{+\frac{7}{2}}_{A} \right) \right. \\
\left. +\beta \left(+\frac{7\sqrt{2}}{8} \ket*{+\frac{5}{2}}_{A} 
+\frac{\sqrt{105}}{8} \ket*{+\frac{1}{2}}_{A} 
-\frac{\sqrt{84}}{8} \ket*{-\frac{3}{2}}_{A} 
-\frac{7}{8} \ket*{-\frac{7}{2}}_{A} \right) \right] \\
\otimes \ket*{-\frac{7}{2}}_{B} \otimes \ket*{-\frac{7}{2}}_{C} \:.
 \label{SX_on_enc_after_disentangle}
\end{split}
\end{equation}

\begin{figure}
\includegraphics[width=8cm]{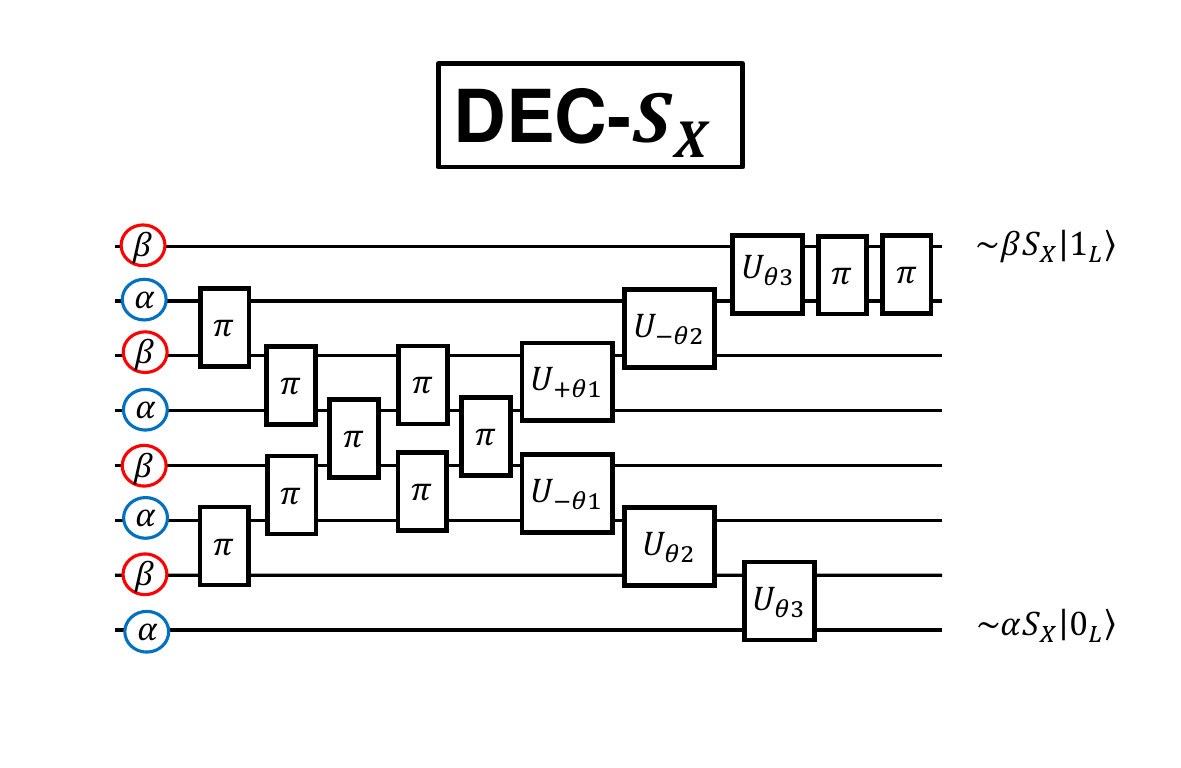}
\caption{ The decoding gates for $S_X$ on qudit A. Unitary rotations $U_{\theta_{i}}$ are around $y$ axis, with cos$(\theta_{i}) = \sqrt{14/29}, \sqrt{12/41}, \sqrt{7/48}$ for $i = 1, 2, 3$, respectively. 
\label{DEC_SX}}
\end{figure}

Then, we apply the DEC($S_X$) operation shown in Fig.~\ref{DEC_SX}. Then our $S_{X,A}$ error state ($\sqrt{\epsilon_{X,A}}(\alpha S_{X,A}\ket{0_L}+\beta  S_{X,A} \ket{1_L})$) will sent to Eqn.~\ref{after_dec_SX}, allowing detection of this error by condition excitation and projective measurement of the ancilla.

\begin{equation}
\begin{split}
\ket{\psi} =  \left( \sqrt{\epsilon_{X,A}}+C \right) \left( \alpha \ket*{+\frac{7}{2}}_{A} +   \beta \ket*{-\frac{7}{2}}_{A} \right) \otimes \ket*{-\frac{7}{2}}_B \otimes \ket*{-\frac{7}{2}}_C \\
+\mathrm{(remaining \, terms)}
\label{after_dec_SX}
\end{split}
\end{equation}

\subsection{The decoding sequence for $S_{X}^2$ error on qudit A case }

Next, we can start the decoding of the $\sqrt{\epsilon_{X^2,A}}(\alpha S_{X,A}^2\ket{0_L}+\beta  S_{X,A}^2 \ket{1_L})$ term. This state can be written as 
\begin{equation}
\begin{split}
\ket{\psi_{S_{X}^2,A}} = c_0 \left[ \alpha \left( +\sqrt{\frac{2}{16}}\cdot\frac{7}{4} \ket*{-\frac{7}{2}}_{A}\ket*{-\frac{7}{2}}_{B,C} +\sqrt{\frac{2}{16}}\cdot\frac{2\sqrt{21}}{4} \ket*{-\frac{3}{2}}_{A}\ket*{-\frac{7}{2}}_{B,C} \right. \right. \\
+\sqrt{\frac{7}{16}}\cdot\frac{2\sqrt{21}}{4} \ket*{-\frac{7}{2}}_{A}\ket*{-\frac{3}{2}}_{B,C} 
+\sqrt{\frac{7}{16}}\cdot\frac{27}{4} \ket*{-\frac{3}{2}}_{A}\ket*{-\frac{3}{2}}_{B,C} \\
+\sqrt{\frac{7}{16}}\cdot\frac{4\sqrt{15}}{4} \ket*{+\frac{1}{2}}_{A}\ket*{-\frac{3}{2}}_{B,C} 
+\sqrt{\frac{7}{16}}\cdot\frac{6\sqrt{5}}{4} \ket*{+\frac{1}{2}}_{A}\ket*{+\frac{5}{2}}_{B,C}\\
\left. +\sqrt{\frac{7}{16}}\cdot\frac{19}{4} \ket*{+\frac{5}{2}}_{A}\ket*{+\frac{5}{2}}_{B,C} \right) \\
+ \beta \left(+\sqrt{\frac{2}{16}}\cdot\frac{7}{4} \ket*{+\frac{7}{2}}_{A}\ket*{+\frac{7}{2}}_{B,C} +\sqrt{\frac{2}{16}}\cdot\frac{2\sqrt{21}}{4} \ket*{+\frac{3}{2}}_{A}\ket*{+\frac{7}{2}}_{B,C} \right.  \\
+\sqrt{\frac{7}{16}}\cdot\frac{2\sqrt{21}}{4} \ket*{+\frac{7}{2}}_{A}\ket*{+\frac{3}{2}}_{B,C}
+\sqrt{\frac{7}{16}}\cdot\frac{27}{4} \ket*{+\frac{3}{2}}_{A}\ket*{+\frac{3}{2}}_{B,C} \\
+\sqrt{\frac{7}{16}}\cdot\frac{4\sqrt{15}}{4} \ket*{-\frac{1}{2}}_{A}\ket*{+\frac{3}{2}}_{B,C}
- \sqrt{\frac{7}{16}}\cdot\frac{6\sqrt{5}}{4} \ket*{-\frac{1}{2}}_{A}\ket*{-\frac{5}{2}}_{B,C}\\
\left. \left. - \sqrt{\frac{7}{16}}\cdot\frac{19}{4} \ket*{-\frac{5}{2}}_{A}\ket*{-\frac{5}{2}}_{B,C} \right) \right] \:,
 \label{SX2_on_enc}
\end{split}
\end{equation}
where $c_0$ stands for a normalization factor, $\sqrt{\frac{8}{357}}$. Then, as discussed in main manuscript, we can divide these into terms parallel and perpendicular to $\ket{0_L}$ (and $\ket{1_L}$).

\begin{equation}
\begin{split}
\ket{\psi_{S_{X}^2,A}} = c_{XX} \left[ \alpha \ket{0_L} + \beta \ket{1_L} \right] \\
 + \left( 1-c_{XX} \right) \left[ \alpha \left( -\sqrt{\frac{2}{16}}\cdot\frac{14}{4} \ket*{-\frac{7}{2}}_{A}\ket*{-\frac{7}{2}}_{B,C} +\sqrt{\frac{2}{16}}\cdot\frac{2\sqrt{21}}{4} \ket*{-\frac{3}{2}}_{A}\ket*{-\frac{7}{2}}_{B,C} \right. \right. \\
+\sqrt{\frac{7}{16}}\cdot\frac{2\sqrt{21}}{4} \ket*{-\frac{7}{2}}_{A}\ket*{-\frac{3}{2}}_{B,C} 
+\sqrt{\frac{7}{16}}\cdot\frac{6}{4} \ket*{-\frac{3}{2}}_{A}\ket*{-\frac{3}{2}}_{B,C} \\
+\sqrt{\frac{7}{16}}\cdot\frac{4\sqrt{15}}{4} \ket*{+\frac{1}{2}}_{A}\ket*{-\frac{3}{2}}_{B,C} 
+\sqrt{\frac{7}{16}}\cdot\frac{6\sqrt{5}}{4} \ket*{+\frac{1}{2}}_{A}\ket*{+\frac{5}{2}}_{B,C}\\
\left. -\sqrt{\frac{7}{16}}\cdot\frac{2}{4} \ket*{+\frac{5}{2}}_{A}\ket*{+\frac{5}{2}}_{B,C} \right) \\
+ \beta \left( -\sqrt{\frac{2}{16}}\cdot\frac{14}{4} \ket*{+\frac{7}{2}}_{A}\ket*{+\frac{7}{2}}_{B,C} +\sqrt{\frac{2}{16}}\cdot\frac{2\sqrt{21}}{4} \ket*{+\frac{3}{2}}_{A}\ket*{+\frac{7}{2}}_{B,C} \right. \\
+\sqrt{\frac{7}{16}}\cdot\frac{2\sqrt{21}}{4} \ket*{+\frac{7}{2}}_{A}\ket*{+\frac{3}{2}}_{B,C}
+\sqrt{\frac{7}{16}}\cdot\frac{6}{4} \ket*{+\frac{3}{2}}_{A}\ket*{+\frac{3}{2}}_{B,C} \\
+\sqrt{\frac{7}{16}}\cdot\frac{4\sqrt{15}}{4} \ket*{-\frac{1}{2}}_{A}\ket*{+\frac{3}{2}}_{B,C}
- \sqrt{\frac{7}{16}}\cdot\frac{6\sqrt{5}}{4} \ket*{-\frac{1}{2}}_{A}\ket*{-\frac{5}{2}}_{B,C}\\
\left. \left. + \sqrt{\frac{7}{16}}\cdot\frac{2}{4} \ket*{-\frac{5}{2}}_{A}\ket*{-\frac{5}{2}}_{B,C} \right) \right] \:,
 \label{SX2_on_enc_divide}
\end{split}
\end{equation}
where $c_{XX}$ stands for $\bra{0_L}S_{X}^2\ket{0_L} = \bra{1_L}S_{X}^2\ket{1_L} = \sqrt{\frac{8}{357}} \cdot \frac{21}{4}$. The second term with coefficient $(1-c_{XX})$ corresponds to $(\alpha \ket{P_{X^2,A,0}} + \beta \ket{P_{X^2,A,1}})$. Since the first term ($c_{XX} [\alpha \ket{0_L} + \beta \ket{1_L}]$) was already projected out during the detection of ``no-error'', detection of the second term only (combination of $\ket{P_{X^2,A,0}} , \ket{P_{X^2,A,1}}$) is sufficient at this step.

\begin{figure}
\includegraphics[width=16cm]{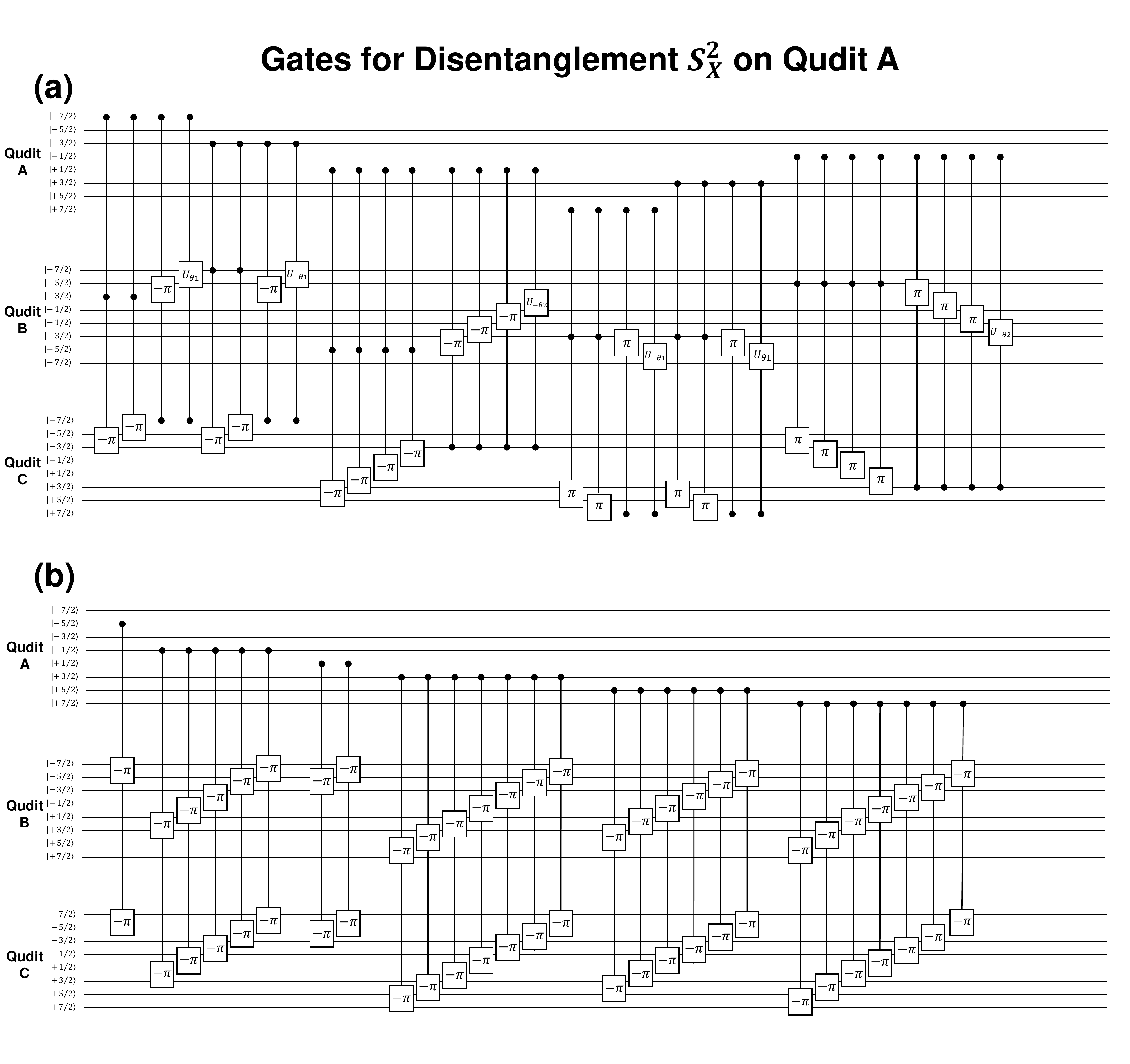}
\caption{ The Disentangle gates for $S_{X}^2$ on qudit A. Unitary rotations $U_{\theta_{i}}$ are around $y$ axis, with cos$(\theta_{i}) = \sqrt{2/5}, \sqrt{4/7}$ for $i = 1, 2$, respectively. Pulses on diagram (a) applied first, and then pulses on (b) applied.
\label{Disentangle_SX2}}
\end{figure}

As above, we can first apply the DISENTANGLE($S_{X}^2$) block, as shown in Fig.~\ref{Disentangle_SX2}.

Following this, the state can be written as
\begin{equation}
\begin{split}
\ket{\psi_{S_X^2,A}} = \left[ \alpha \left( -\frac{14\sqrt{5}}{64} \ket*{-\frac{7}{2}}_{A} 
+\frac{2\sqrt{105}}{64} \ket*{-\frac{3}{2}}_{A} 
+\frac{14\sqrt{15}}{64} \ket*{+\frac{1}{2}}_{A} 
-\frac{2\sqrt{7}}{64} \ket*{+\frac{5}{2}}_{A} \right) \right. \\
\left. +\beta \left(-\frac{14\sqrt{5}}{64} \ket*{+\frac{7}{2}}_{A} 
+\frac{2\sqrt{105}}{64} \ket*{+\frac{3}{2}}_{A} 
+\frac{14\sqrt{15}}{64} \ket*{-\frac{1}{2}}_{A} 
+\frac{2\sqrt{7}}{64} \ket*{-\frac{5}{2}}_{A} \right) \right] \\
\otimes \ket*{-\frac{7}{2}}_{B} \otimes \ket*{-\frac{7}{2}}_{C} \:.
 \label{SX2_on_enc_after_disentangle}
\end{split}
\end{equation}

\begin{figure}
\includegraphics[width=8cm]{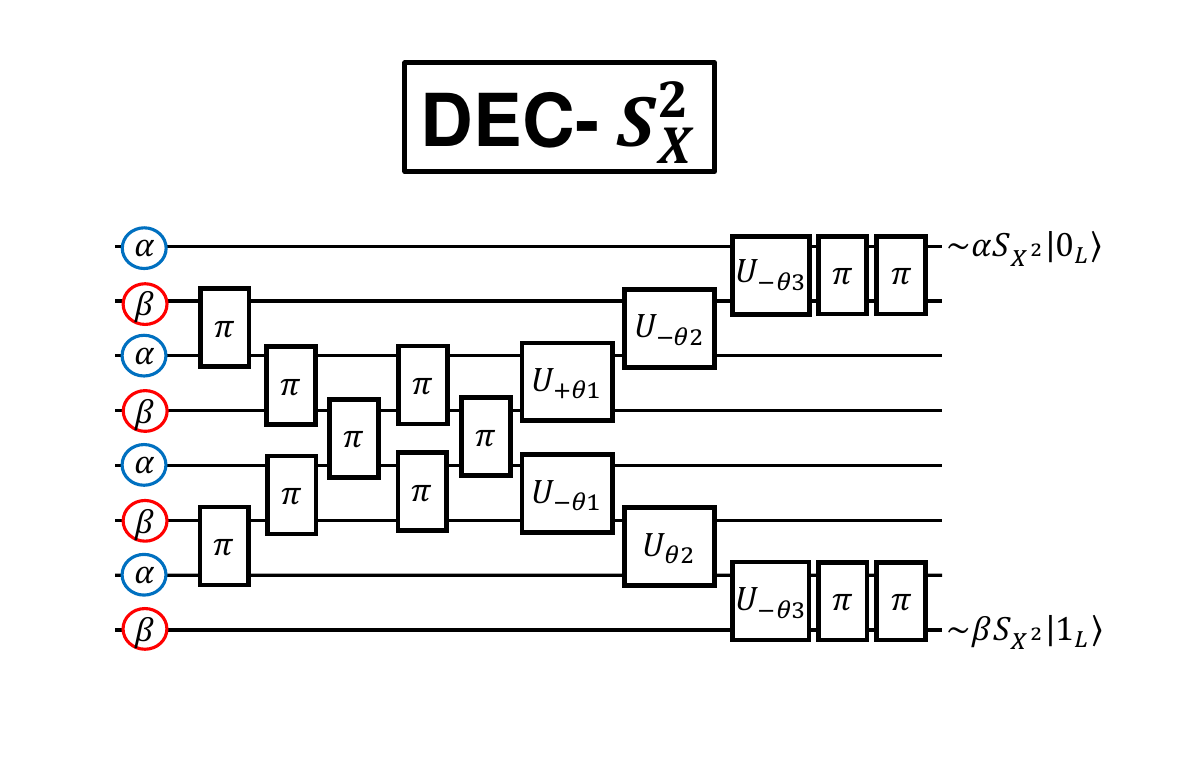}
\caption{ The decoding gates for $S_X^2$ on qudit A. Unitary rotations $U_{\theta_{i}}$ are around $y$ axis, with cos$(\theta_{i}) = \sqrt{1/106}, \sqrt{15/121}, \sqrt{35/156}$ for $i = 1, 2, 3$, respectively. 
\label{DEC_SX2}}
\end{figure}

Finally, the DEC($S_{X}^2$) operation, shown in Fig.~\ref{DEC_SX2}, is applied. This sends the $S_{X,A}^2$ error state ($\sqrt{\epsilon_{X^2,A}}(\alpha S_{X,A}^2\ket{0_L}+\beta  S_{X,A}^2 \ket{1_L})$) to the state 
\begin{equation}
\begin{split}
\ket{\psi} = \left( \sqrt{\epsilon_{X^2,A}}+C \right) \left( \alpha \ket*{-\frac{7}{2}}_{A} +   \beta \ket*{+\frac{7}{2}}_{A} \right) \otimes \ket*{-\frac{7}{2}}_B \otimes \ket*{-\frac{7}{2}}_C \\
+\mathrm{(remaining \, terms)} \:,
\label{after_dec_SX2}
\end{split}
\end{equation}
and, by analogy with above, conditional excitation and projective measurement of the ancilla detects this error case.

\subsection{The decoding sequence for other errors }

\begin{figure}
\includegraphics[width=16cm]{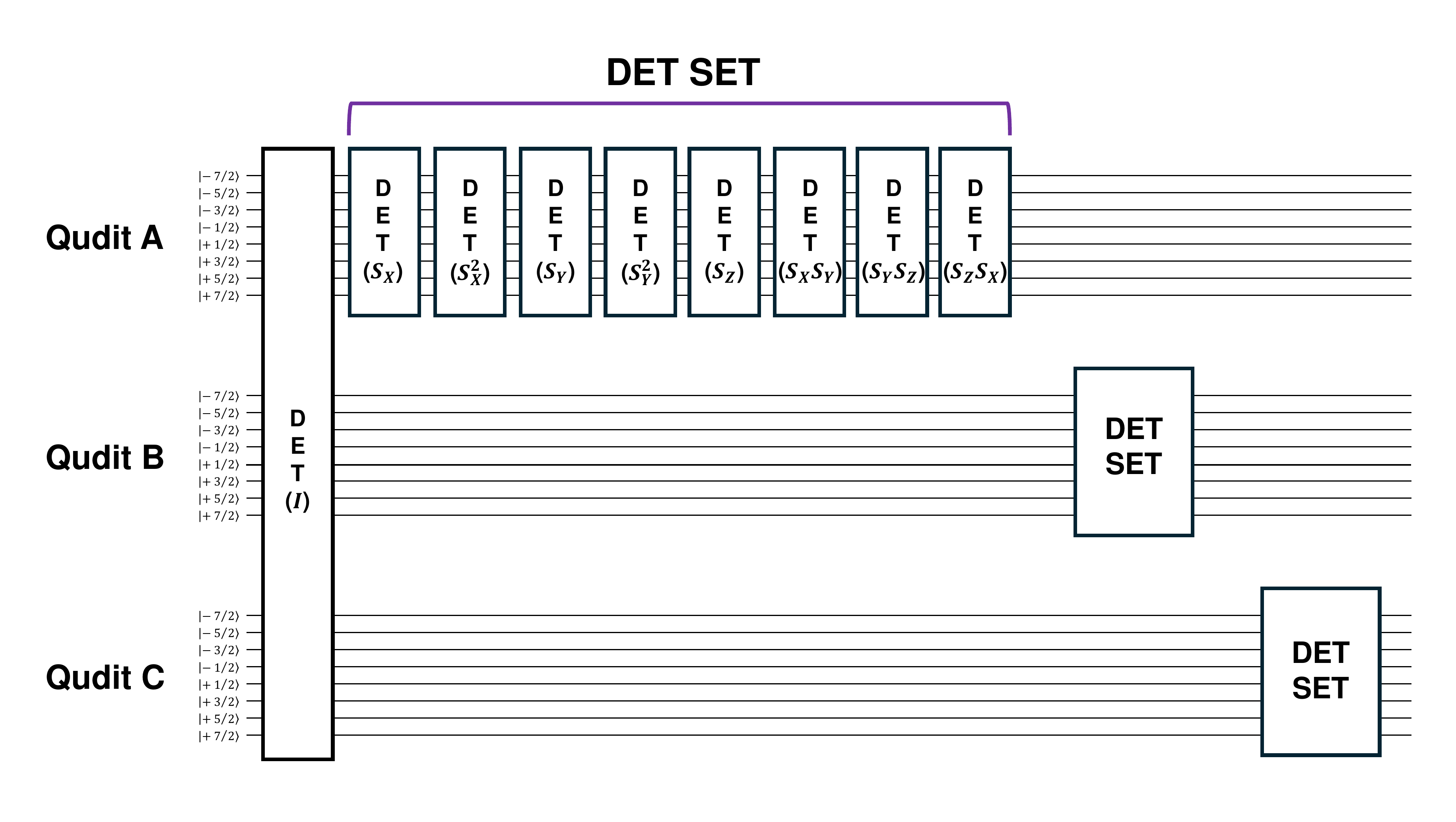}
\caption{ The schematic diagram of full decoding pulse sequence for multi-qudit error correction code. Each DET code block including Disentanglement -- decoding -- conditional ancilla measure -- recovery sequence for each error case.
\label{Full-dec}}
\end{figure}

Fig.~\ref{Full-dec} shows full the decoding sequence blocks for the first-, second- and third qudit of multi-qudit-error-correction sequence. Each `DET' code block includes the sequence:  Disentanglement -- decoding -- conditional ancilla measurement  -- recovery. Depending on the result, the ancilla qubit can be initialized and original information can be extracted. After full decoding and detection of the first qudit error case, as shown in Fig.~\ref{Full-dec}, the error on the other two qudits can be decoded and detected by repeating the analogous procedure for each qudit in turn.

Here, we have only provided detailed pulses composing the blocks ENC, DET-I, DET-S$_X$, DET-S$_{X}^2$, but the sequences for the DET blocks can be obtained by calculating the unitary transformation between each error case and the state in Eqn.~\ref{State_after_error}. It is worth noting again that with each step of the detection scheme, we project out components of the error state in sequence.

\end{document}